\documentclass[manuscript]{aastex}

\usepackage{multirow}

\slugcomment{}

\shorttitle{First spectral optical monitoring of E1821+643}
\shortauthors{Shapovalova et al.}

\begin{document}

%% LaTeX will automatically break titles if they run longer than
%% one line. However, you may use \\ to force a line break if
%% you desire.

\title{First long-term optical spectro-photometric monitoring of a binary black hole 
candidate E1821+643: I. Variability of spectral lines and continuum}

%% Use \author, \affil, and the \and command to format
%% author and affiliation information.
%% Note that \email has replaced the old \authoremail command
%% from AASTeX v4.0. You can use \email to mark an email address
%% anywhere in the paper, not just in the front matter.
%% As in the title, use \\ to force line breaks.

\author{A.I. Shapovalova\altaffilmark{1}, L.\v C. Popovi\'c\altaffilmark{2,3,4}, 
V.H. Chavushyan\altaffilmark{5}, A.N. Burenkov\altaffilmark{1}, D. Ili\'c\altaffilmark{3,4}, \\
W. Kollatschny\altaffilmark{6}, A. Kova\v cevi\'c\altaffilmark{3,4}, J.R. Vald\'es\altaffilmark{5}, 
V. Pati\~no-\'Alvarez\altaffilmark{5},\\
J. Le\'on-Tavares\altaffilmark{5}, J. Torrealba\altaffilmark{5} and V. E. Zhdanova\altaffilmark{1}}

\email{ashap@sao.ru}

\altaffiltext{1}{Special Astrophysical Observatory of the Russian AS,
Nizhnij Arkhyz, Karachaevo-Cherkesia 369167, Russia}
\altaffiltext{2}{Astronomical Observatory, Volgina 7, 11160 Belgrade 74, Serbia}
\altaffiltext{3}{Department of Astronomy, Faculty of Mathematics, University
of Belgrade, Studentski trg 16, 11000 Belgrade, Serbia}
\altaffiltext{4}{Isaac Newton Institute of Chile, Yugoslavia Branch, Volgina 7, Belgrade, Serbia}
\altaffiltext{5}{Instituto Nacional de Astrof\'{\i}sica, \'{O}ptica y
Electr\'onica, Apartado Postal 51-216, 72000 Puebla, Puebla, M\'exico}
\altaffiltext{6}{Institut f\"ur Astrophysik, Georg-August-Universit\"at G\"ottingen, Germany}

%% Mark off your abstract in the ``abstract'' environment. In the manuscript
%% style, abstract will output a Received/Accepted line after the
%% title and affiliation information. No date will appear since the author
%% does not have this information. The dates will be filled in by the
%% editorial office after submission.

\begin{abstract}

{ We report the results of the first long-term (1990--2014) optical 
spectro-photometric monitoring of a binary black hole candidate QSO E1821+643, a low-redshift
high-luminosity radio-quiet quasar. In the monitored period
the continua and H$\gamma$ fluxes changed for around two times, while the
H$\beta$ flux changed around 1.4 times. We found the
periodical variations in the photometric flux with the periods of 1200, 
1850 and 4000 days, and 4500 days periodicity in the spectroscopic  variations.
However, the periodicity of 4000--4500 days covers only one cycle of
variation and should be confirmed with a longer monitoring campaign. 
There is an indication of the period around 1300 days in the spectroscopic light 
curves, but with small significance level, while the 1850 days period could not be
clearly identified in the spectroscopic light curves.

The line profiles have not significantly changed, showing an important red 
asymmetry and broad line peak redshifted around +1000 km s$^{-1}$. However, H$\beta$ 
shows  broader mean profile and has
a larger time-lag ($\tau\sim120$ days) than H$\gamma$ ($\tau\sim60$ days). 
We estimate that the mass of the
black hole is $\sim2.6\times10^9\rm M_\odot$.

The obtained results are discussed in the frame of the binary black hole
hypothesis. To explain the periodicity in the flux variability and high
redshift of broad lines we discuss a scenario where dense gas-rich
cloudy-like structures are orbiting around a recoiling black hole.}

\end{abstract}

%% Keywords should appear after the \end{abstract} command. The uncommented
%% example has been keyed in ApJ style. See the instructions to authors
%% for the journal to which you are submitting your paper to determine
%% what keyword punctuation is appropriate.


\section{Introduction}

Remarkable features in the optical spectrum of quasars
are very broad emission lines which very often show variability in 
the flux and shape. With the monitoring of these lines one can investigate
the structure of the emitting gas \citep[i.e. the broad-line region - BLR, see e.g.][]{su00,sh10,pop11},
the mass of the supermassive black hole (SMBH) that resides
in the center of quasars \citep[see e.g.][]{pe14}, and study the 
signatures of galactic evolution i.e. black hole merger effects \citep[see e.g.][]{bo12,pop12,bo15}. 

QSO E1821+643 is one of the most luminous radio-quiet quasar in 
the local universe (z=0.297,  $m_{\rm V}=14.2$, $M_{\rm V}=-27.1$),
first detected as an unusually very strong soft X-ray emitter \citep[][]{pm84,ru10}.
The multiwavelenght observations (from the X-ray to infrared) by \cite{ko93}
gave the spectral energy distribution following a power-low ($\alpha=1.16$) mainly in the IR and X-ray, while the
strong optical/UV "blue bump" was modeled with the thermal accretion disk, 
yielding the mass of the central SMBH of $3 \times 10^9 M_{\odot}$, and an accretion rate of $19 M_{\odot}/{\rm year}$
\citep[][]{ko93}.
E1821+643 exhibits the remarkable UV/optical/IR emission-line spectrum \citep[][]{ko91,ko06,la08}
with very broad emission lines (line widths $>$5000 kms$^{-1}$), where also absorption lines
are seen in the UV (Ly$\alpha$, CIV, and OIV lines), that may be due to absorption by the gas associated with the quasar 
\citep[][]{ba92,oe00}.

Even though it has many features usually seen in radio-loud quasars (a high luminosity,
an elliptical host galaxy, surrounding cluster of Abell richness class $>$2), this object is { marked}
as a radio-quiet { based on} its radio and nuclear [OIII] line luminosity \citep[][]{la92}. 
The milliarcsecond-resolution radio images { of this radio-quiet quasar} showed
that { its compact radio emission is produced by a black hole based jet, 
rather than a starburst} \citep[][]{bl95,bl96}, { while} the deep VLA observations revealed 
a radio emission over $280 h^{-1}$ kpc, { extended far} beyond the host galaxy \citep[][]{bl01}.

The object is embedded in a large elliptical galaxy $M_{\rm gal} \approx 
2 \times 10^{12} M_{\odot}$ \citep[][]{mm01,fl04}
that is associated with the rich cluster of galaxies \citep[][]{hn91},
and most likely it resides in its center \citep[][]{sc92}. There are some indications that the 
quasar SMBH interacts with the surrounding intracluster medium \citep[see in more details][]{oe00,re14,wa14}.

The off-nuclear optical spectrum has shown the extended emission-line gas, with the [OIII] luminosity $\sim$2 orders 
of magnitude higher than in other radio-quiet QSOs \citep[][]{fr98}.
This extra-nuclear gas is probably due to tidal interactions or merger processes \citep{fr98}. 
Also \cite{ar11} observed $^{12}$CO J = 1-0 emission line in E1821+643 and 
found that the CO emission is likely to be extended, showing a high asymmetry
with respect to the center of the host elliptical where the QSO resides. This also suggests that the CO emission may be connected with merger effects, i.e. {may come from a gas-rich companion galaxy in merger or may be a tail-like structure from a previous interaction} \citep{ar11}. 
Moreover, the broad emission Balmer lines show an unusual shape. They have highly
red asymmetric profiles and at the same time are redshifted ($\approx$1000 km s$^{-1}$) relative to the narrow lines 
\citep[][]{la08,ro10}, which may be caused by the emission of one active component of a binary SMBH or recoiling black hole after SMBH collision \citep[see][]{pop12}. \cite{ro10} analyzed spectropolarimetric observations of E1821+643 and found that
the central SMBH is itself moving with a velocity $\sim$2100 km s$^{-1}$ relative to the host galaxy, that indicates  
a gravitational recoil that follows the merger of a SMBH binary system.

Here we present the long-term (1990--2014) optical { spectro-photometric} monitoring of QSO E1821+643, that
is the first monitoring campaign of this object. The motivations for long-term observations were (as a summary of discussion above) that E1821+643: i) is very bright quasar located in the (center of) galaxy cluster; ii) is hosted in an elliptical galaxy, but is a radio-quite object; iii) has a SMBH that probably interacts with the surrounding intracluster medium, and iv) is probably experiencing a black hole merger (reminiscent of a previous black hole merger or current supermassive binary black hole interaction). All these facts pay attention to  E1821+643 to be monitored with the aim to find some specific behaviors in the optical spectral variation
(in the continuum and line fluxes), and to investigate the nature of this object.

In this paper (Paper I) we present photometric and spectroscopic observations of QSO E1821+643,  analyzing variability in the H$\gamma$, H$\beta$ line and continuum fluxes. In the forthcoming paper (Paper II) we will 
 give  detailed analysis of the broad line parameters, spectral energy distribution and Balmer continuum variabilities.
The { structure of this paper is as follows}: in section 2 we describe the observations and reduction of observed data, in section 3 we analyze the observed spectral data and give results, in section 4 the obtained results are discussed, and in section 5 we { present} our main conclusions.

\section{Observations and data reduction}

\subsection{Photometry}

\begin{figure*}
\centering
\includegraphics[height=17cm, angle=270]{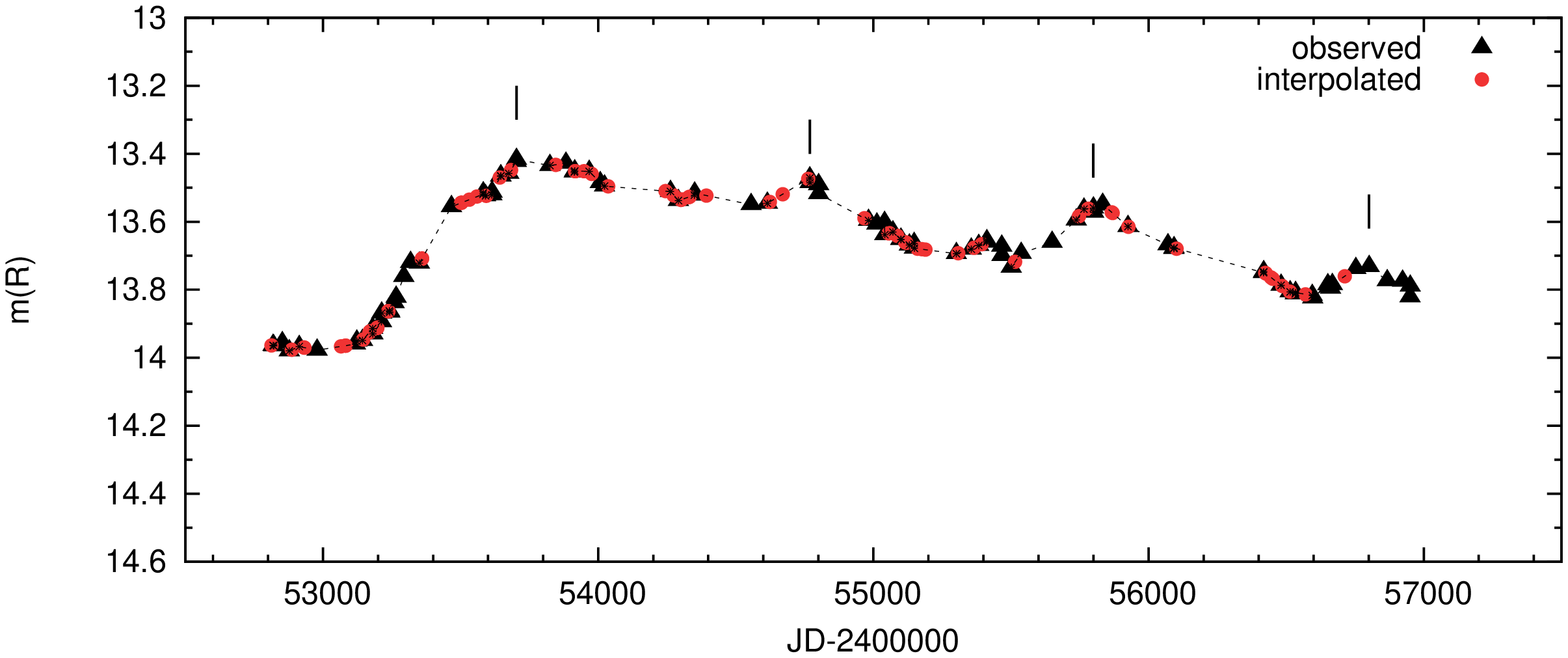}
\caption{R-band light curve for observed and interpolated data. The four maxima are denoted on the plot.
 The error bars are comparable to the symbol size.} \label{fig1}
\end{figure*}

{ The photometry in BVR filters of E1821+643 was performed at the Special Astrophysical Observatory of the Russian Academy of Science (SAO RAS) during the 2003--2014 period (98 nights) with the 1m Zeiss telescope with an offset guided automatic photometer. The photometer contains a CCD camera of the size of 1040$\times$1160 pixels, that is cooled with liquid nitrogen \citep{am00}. The pixel scale at the CCD is 0.45\arcsec /pixel, that correspons to 7.5$\times$8.5 arcmin field of view.
Both bias and dark current frames were taken, while for the flat-filed frames we adopted the morning and evening sky exposures. The software developed at SAO RAS by \citet{vl93} was used for the data reduction. 
The photometry is done by integrating the signal in the concentric circular apertures of increasing size,
that are centered at the baricenter of the measured object. The photometric system of this instrument 
resemble those of Johnson in B and V filter, and of Cousins in R filter \citep{co76}. For local photometric standards we used stars of \citet{pen71} that are close to the position of E1821+643 on our CCD images, which results with the negligible effects of differential air mass.} In Table \ref{tab1} (available electronically only) the photometric BVR-magnitude for the aperture of 15\arcsec are presented and in Fig. \ref{fig1} we plotted the light curve in the R-band.

\subsection{Spectral observations}

Spectra of E1821+643 ($\sim$140 nights) were { acquired with two telescopes (6 m and 1 m) of the 
SAO RAS, Russia (during 1998--2014), one telescope (INAOE's 2.1 m) of the Guillermo Haro Observatory (GHO) 
Cananea, Sonora, M\'exico (during 1998--2007, and 2013), and two telescopes (3.5 m and 2.2 m) of Calar Alto Observatory (CAO), Spain (during 1990--1994). All spectra were acquired with long-slit spectrographs with CCDs. 
The representative wavelength interval was from 4000 \AA \ to 7500 \AA, with the spectral resolution between 
4.5 \AA \ and 15 \AA, and the S/N ratio $>$50 in the continuum close to the H$\beta$ line. Every night the spectrophotometric standard stars were observed. Information on the source of spectral observations is listed in Table \ref{tab2}, while the log of spectral observations is listed} in Table \ref{tab3} (available electronically only).

The spectrophotometric data reduction was { performed using the software developed at SAO RAS, while for 
the spectra acquired in Mexico and in Spain the IRAF package was used. The image reduction procedure consisted from the standard bias and flat-field corrections, cosmic ray removal, 2D wavelength linearization, sky spectrum subtraction, addition of the spectra for every night, and relative flux calibration based on standard star. Further in the analysis, we rejected approximately 10\% of spectra due to different reasons (e.g. poor spectral resolution ($>$15 \AA), high noise level, badly corrected spectral sensitivity, etc.), thus our final data set 
contains 127 spectra, that are further analyzed.}

\subsection{Absolute calibration (scaling) of the spectra}

The { common} technique of { the spectral flux calibration based on the comparison with stars of determined spectral energy distribution, is not precise enough for the study of an AGN variability, as the spectrophotometric accuracy is not less than 10\% even under great photometric conditions. Thus the standard stars were only used for
a relative flux calibration. For the absolute calibration, the AGN narrow emission lines fluxes were used for scaling the spectra, because it is noted that these lines do not vary on time scales of decades \citep{pet93}.}

E1821+643 is a QSO with very bright [OIII] $\lambda\lambda$4959,5007 emission lines and it 
is possible to scale our spectra using the flux of these lines.  However, a problem may be that the [OIII] 
lines are varying during the long monitored period of 24 years. 

The very bright [OIII] emission \citep[two orders of magnitude brighter than in other radio-quit quasars, see][]{fr98} indicates the presence of a very large narrow line region (NLR). Taking the relation $R_{\rm NLR}\sim L[{\rm OIII}]^{0.33}$ given by  \cite{sc03} and the [OIII]$\lambda\lambda$4959, 5007 flux of the order of $\sim 2\times 10^{-13} \rm{erg \, cm^{-2} s^{-1}} $ for E1821+643 (z$\approx0.3$), we estimated that the size of the extended NLR is of the order of $\sim$8 kpc (26000 light years), that is three orders of magnitude larger than our monitored period (24 years), consequently one cannot expect to detect the variability in the [OIII] line flux during this monitored period. 

%One should bare in mind that this extended [OIII] emission is comparatively faint and the unresolved NLR still dominates the integrated [OIII] flux. However, it is reasonable to assume that it does not change significantly over timescales less than a few years, since even the unresolved NLR will have a light crossing time of at least a few decades. Therefore, we suppose that the [OIII] variations are not responsible for the observed variability, which typically occurs on a timescale of $\sim$ few 100 days.

Therefore, we used the  [OIII]4959+5007 integrated line flux, that is taken to be $2.9\times 10^{-13} \rm{erg \, cm^{-2} s^{-1}} $ \citep[][]{ko93}, in order to absolutely calibrate our spectra { with the method proposed by} \citet{vg92} { that is} modified by \citet{sh04}. Note that this method was tested for different S/N$>$20 and different spectral resolutions, and it has been shown that the uncertainty in the scaling is $\sim$1-2\%, and 
depends only on the quality of the spectra. For more details see \citet{vg92} and Appendix A in \citet{sh04}.

%\footnote{see For more details see Appendix A in \citet{sh04}}. 

\subsection{Unification of the spectral data}

{ For the study of the long term spectral variability of an AGN
observed with telescopes of different apertures, it is mandatory to construct 
an uniformed data set in a consistent way. Since instruments of different apertures were used for the observations, it is necessary to correct for aperture effects both the continuum and line fluxes \citep{pc83}. 
Based on our past work \citep{sh01, sh04, sh10, sh12, sh13}, we determined a point-source scale correction factor 
($\varphi$) and an aperture-dependent correction factor that corrects for the host galaxy contribution to the continuum, i.e. extended-source correction factor (G(g)). We} used the following expressions (see \citet{pet95}):
$$F({\rm line}){\rm true} = \varphi * F({\rm line}){\rm obs} ,$$
$$ F({\rm cont}){\rm true} = \varphi * F({\rm cont}){\rm obs} - G(g), $$
where index "obs" denotes the observed flux, and "true" the aperture corrected flux.
The spectra of the 6m telescope, within an aperture of
2\arcsec $\times$ 6\arcsec were adopted as standard (i.e. $\varphi$= 1.0, G(g)=0 by definition). 
{ The correction factors $\varphi$ and G(g) are estimated empirically, comparing the observations from all telescope data sets with the simultaneous one from the standard data set (the same method was used in AGN Watch, see e.g. \citet{pet94,pet98,pet02}). Intervals which we noted as "nearly simultaneous" are practically of
1-3 days, suppressing the variability on short time scales ($<$ 3 days).
The point-source scale correction factor $\varphi$ and extended-source correction factor 
G(g) values (in units of 10$^{-15} \rm erg \, cm^{-2} s^{-1} \AA^{-1}$) are given for different data sets in Table \ref{tab4}.}

\subsection{Measurements of the spectral fluxes and errors}

From the scaled spectra (see sections 2.2--2.3)  { we measured the continuum flux}  near the H$\beta$ 
line at the observed wavelengths $\sim 6616$ \AA\, ($\sim 5100$ \AA\, in the rest frame), by averaging fluxes in the spectral range of 6601--6631 \AA\, and { the continuum flux} near the H$\gamma$ line at the observed wavelengths $\sim 5474$ \AA\, ($\sim 4220$ \AA\, in the rest frame), by averaging  fluxes in the spectral range of 5459--5489 \AA. Note that the spectrum of E1821+643  does not contain significant absorption lines in the observed spectral range (see Fig. \ref{example}).

{ For the determination of the H$\beta$ and H$\gamma$ fluxes, we must first subtract the
underlying continuum. For this goal, a linear continuum was fitted through the windows of 20\,\AA\,,
6160--6635\,\AA\, for the H$\beta$ region, and 5480--5780\,\AA\, 
for the H$\gamma$ region. Then, the observed line fluxes were measured in the 
following wavelength intervals: (6170--6620)\,\AA\, for H$\beta$  and (5550--5770)\,\AA\, for H$\gamma$.}
Using $\varphi$ and G(g) factors from Table \ref{tab4}, we re-calibrated the observed 
H$\gamma$ and H$\beta$ fluxes, and their corresponding near-by continuum fluxes to a common scale using the standard aperture of 2.0\arcsec$\times$6.0\arcsec.
In Table \ref{tab5} (available electronically only) the fluxes for the continuum at the rest frame wavelength of 5100 \AA\, and 4220 \AA , the total H$\gamma$ and H$\beta$ lines, and their errors are given. 
The listed total H$\gamma$ and H$\beta$ fluxes include the contribution of the narrow H$\gamma$ and H$\beta$ component and the [OIII] 4363, 4959, 5007 lines.
The mean errors (uncertainties) of the 
continuum fluxes at 5100 \AA\, and 4220 \AA\, H$\beta$, and H$\gamma$ lines are 
$\sim$(2-3.5)\%, $\sim$3.5\%, and $\sim$5\%, respectively (see Table \ref{tab6}). 
These quantities were estimated by comparing results from spectra obtained within 
a time interval shorter than 3 days.  Note that the errors in fluxes  in Table \ref{tab5}  were obtained using the mean error from Table \ref{tab6}. Figure \ref{err} shows mean errors in the flux measurements as function of the corresponding mean line fluxes for H$\beta$ and H$\gamma$, and for the continuum at 5100\AA\, and 4220\AA.  The correlation coefficient 
and corresponding p-value are also given in Figure \ref{err}. As can be seen it is obvious that the dependence between 
the mean error and the mean fluxes does not exist between F(H$\beta$) and F(5100), or have a very 
weak trend (statistically insignificant) between F(H$\gamma$) and F(4200).

The contribution of the host galaxy can be very important in some objects \citep[see e.g.][]{sh13},
especially if the stellar absorption lines are present in the AGN spectrum.
Therefore we roughly estimated the contribution of the host 
galaxy from the aperture photometry with apertures 10\arcsec and 15\arcsec. One can expect that 
in the ring between 15\arcsec and 10\arcsec, the flux in V and R filters is from 
the host galaxy, therefore we used the ratio of the ring flux to the 10\arcsec -aperture flux 
to estimate the host-galaxy contribution. We obtained that the contribution of the host galaxy 
in the V-band is from 3.0\% to 4.6\%, relative to the flux in the maximum and minimum activity 
states of the nucleus, respectively. In the R-band the host galaxy contribution is from $\sim$3\% to 9\%.
This is not in contradiction with an estimate made in the V-band by \cite{fl04}
from the modeling of an HST/WFPC2 image of this object. They found that the host-galaxy
contributes around $\sim$10\% to the total luminosity in the V-band in 2000 (when the object 
was close to the minimum state). Also, we should note that in the spectra of E1821+643 
the stellar absorption lines such as Mg Ib (5170), Ca II (H,K), and G-band (4302) are absent, 
which indicates a very small host-galaxy contribution.

\subsection{The broad H$\beta$ line and its segment fluxes}

In this paper we will extract only the broad component of the H$\beta$ line and study the behavior of its 
line segments  during the  monitored period.  The detailed analysis of the H$\beta$ and H$\gamma$ line  
profiles using the Gaussian-fitting method will be given in the forthcoming paper (Paper II).

To obtain only the broad component of the H$\beta$ line we subtracted the underlying continuum that
was fitted through the continuum windows using the B-spline python routine {\texttt interpolate.splev}. 
Then we applied an automatic and self-consistent multi-Gaussian fitting method for removing the narrow 
components of H$\beta$, [OIII]$\lambda\lambda$4959,5007\AA\, lines, and  Fe II lines \citep{pop04,ko10,sh12}
The fitting-method is based on the 
$\chi^2$-minimalization routine, and we reduced  as many Gaussian parameters as possible: i) all narrow lines 
have the same widths and shifts \citep[see][]{pop04}; ii) the flux ratio of [OIII]$\lambda$4959\AA\, and [OIII]$\lambda$5007\AA\, is 1:3 
\citep[][]{dim07}; iii) two broad Gaussian functions (broad and very broad line component) were used to fit the broad H$\beta$ component; iv) the ionized iron line multiplets have lines of same widths and shifts \citep{ko10,sh12}. 
One example of the best fit (thick solid line) of the observed spectrum (dots) is presented in upper panel of Figure \ref{gauss}, where the 42 Fe II multiplet (dashed line) is also shown. The residual (thin solid line), 
and the Gaussian broad and narrow components (dashed lines) are shifted below for better visibility 
(Fig. \ref{gauss}). Note here that only one Gaussian cannot properly fit the narrow line wings of strong [OIII] lines, therefore, we included an additional, broader, Gaussian (with significantly smaller intensity) to fit the narrow [OIII] lines wings.

We determined the flux of the broad H$\beta$ component in the same wavelength range as for the total 
H$\beta$ line and estimated that the contribution of the narrow components is $\sim$30\%. 
We also found that the peak of the broad  H$\beta$ component is shifted by $\sim$1000 km s$^{-1}$ relative to the peak of the narrow  component.  Measuring the flux of the broad H$\beta$ line-segments we divided the broad  
H$\beta$ component relative to the shifted  center (4877.5 \AA, in the rest-frame) into four line-segments with 
the width of $\sim$4000 km s$^{-1}$ each: blue wing, core, red wing, and far red wing (see the bottom panel in Figure \ref{gauss} and Table \ref{tab6}). 
Then we determined the fluxes of these line-segments (Table \ref{segm}, available electronically only). 
The mean errors for the broad H$\beta$ line-segment fluxes are about 3\%-6\% (Table \ref{tab6}).  
As expected, the maximal error  $\sim$6\%  corresponds to the distant, far-red wing and is caused by a bad subtraction of the bright [OIII]$\lambda$5007\AA\, line, and a weak line intensity in this spectral region. 
The mean errors (in \%) of fluxes and the line-segment fluxes are given in Table \ref{tab6} and 
in Table \ref{segm} (available electronically only), respectively.

\section{Data analysis and results}

\subsection{Photometric results}

In Table \ref{tab1} the results of the broad-band photometry of E1821+643 in BVR filters for a circular aperture of 15\arcsec and the corresponding errors are listed.  A light curve in the R-band shows an almost sinusoidal change with the maximum amplitude of about 0.5 magnitude (see Figure \ref{fig1}). There, we can see several ($\sim$4) noticeable peaks (flares, outburst) with different brightness amplitudes. In the BV-bands the light curves have the same shape. Some information about these {flare-like events} are listed in Table \ref{flare}. It follows that the amplitude of  peaks varies from 0.1 to 0.54 magnitude, or in intensity from 1.1 to 1.6 times. It is very interesting to note that the difference in days between the two consecutive flare-like events is $\sim$1000 days (Table \ref{flare}). This indicates some periodicity in the flux variability during the monitored period. Further
(in section 3.2.4) we study the periodicity of light curves using different methods.

\subsection{Spectral results} 

\begin{figure}
\centering
\includegraphics[width=13cm]{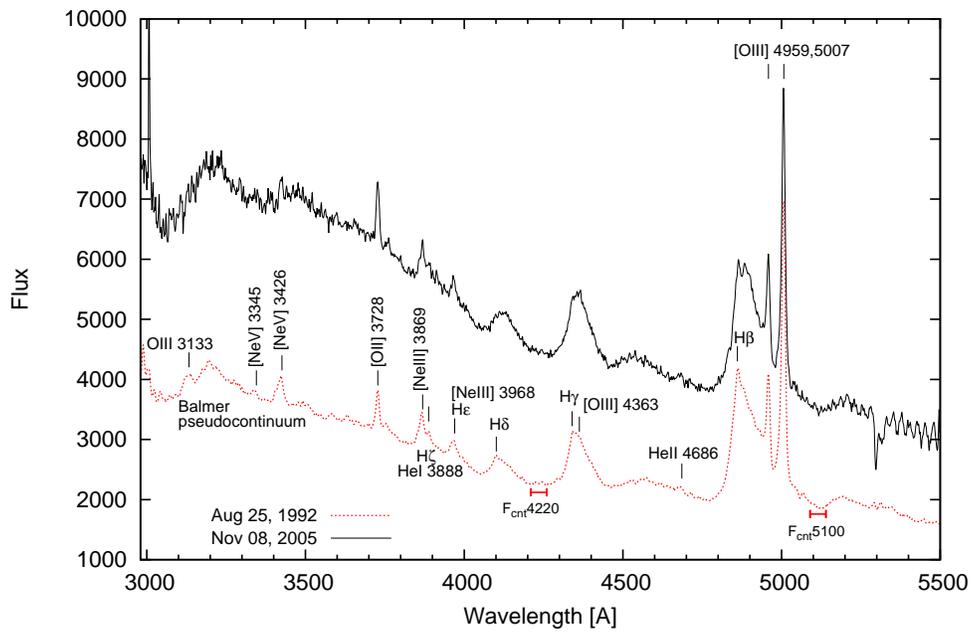}
\caption{Example of two E1821+643 rest-frame spectra observed close to the minimum (down) and maximum (up) of activity in the 
monitoring period. The observed lines and  continuum windows are denoted on the plot.} \label{example}
\end{figure}

\begin{figure*}
\centering
\includegraphics[width=8cm]{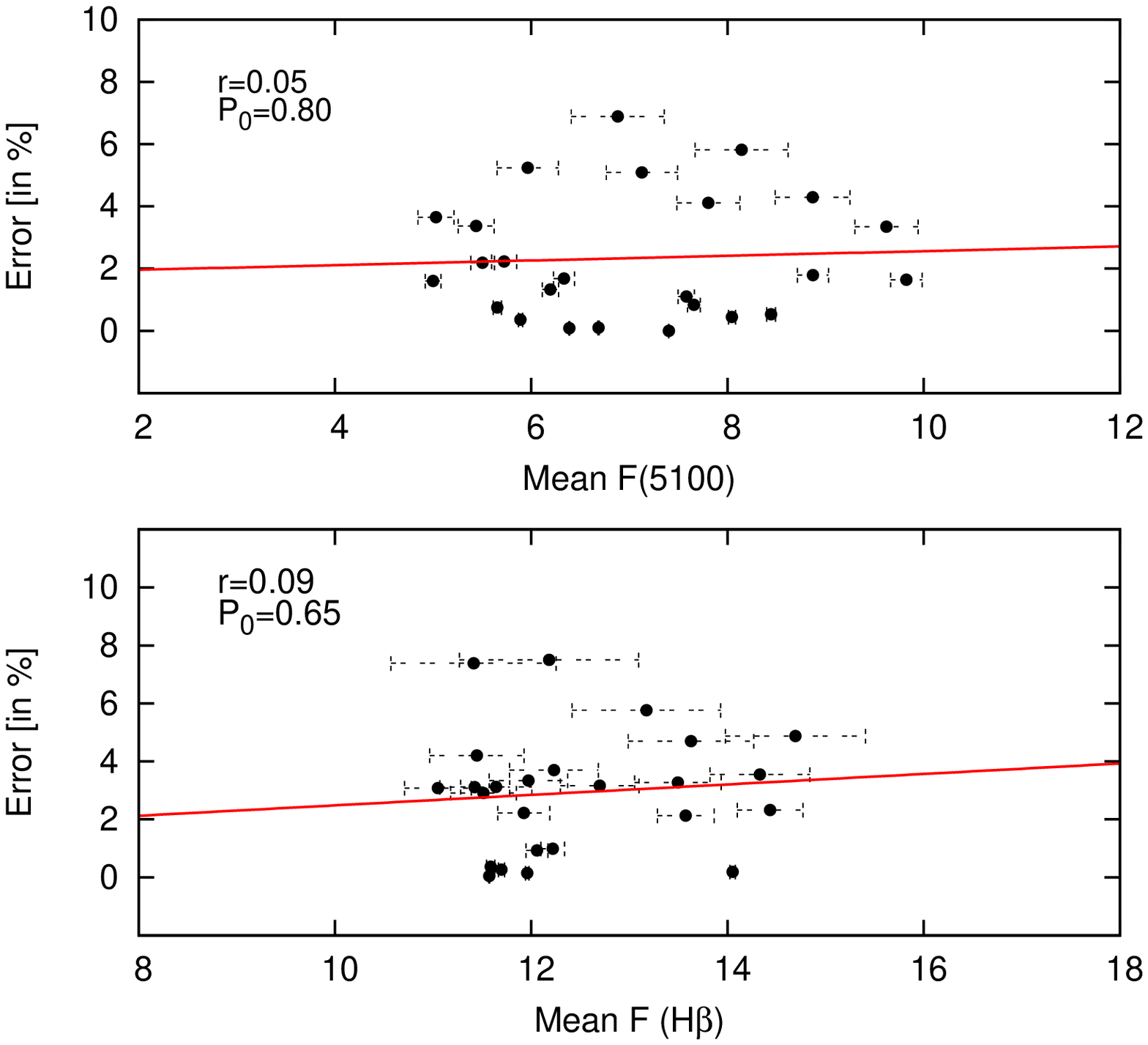}
\includegraphics[width=8cm]{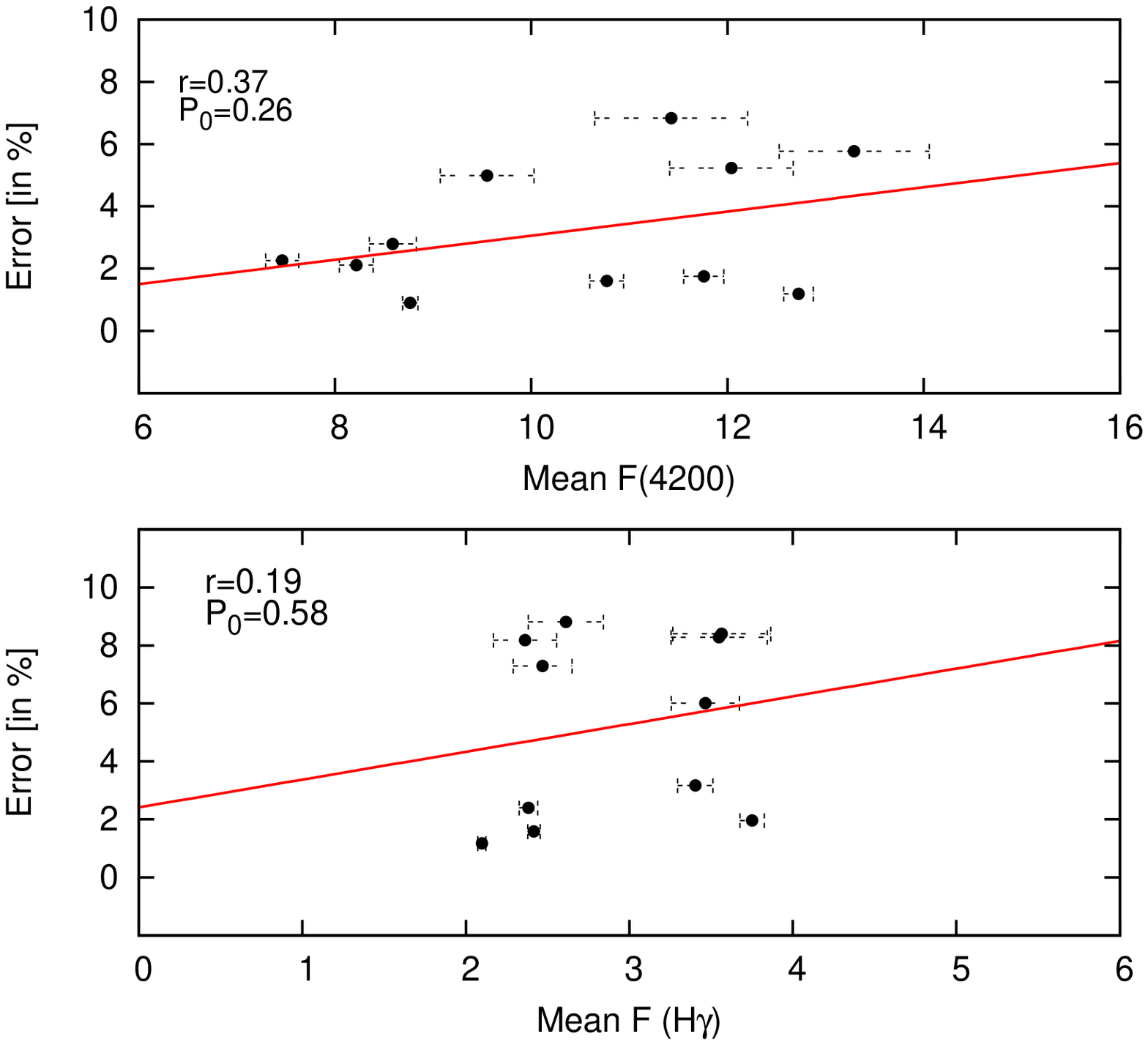}
\caption{The H$\beta$ line and continuum at 5100 \AA\, flux errors (left panels), 
and H$\gamma$ line and continuum at 4200 \AA\, flux errors (right panels) with respect to
the corresponding mean flux. The continuum flux is in units of
10$^{-15}$ erg cm$^{-2}$s$^{-1}$\AA$^{-1}$, and the line fluxes are in units of
10$^{-13}$ erg cm$^{-2}$s$^{-1}$.} \label{err}
\end{figure*}

\begin{figure}
\centering
\includegraphics[width=10.5cm]{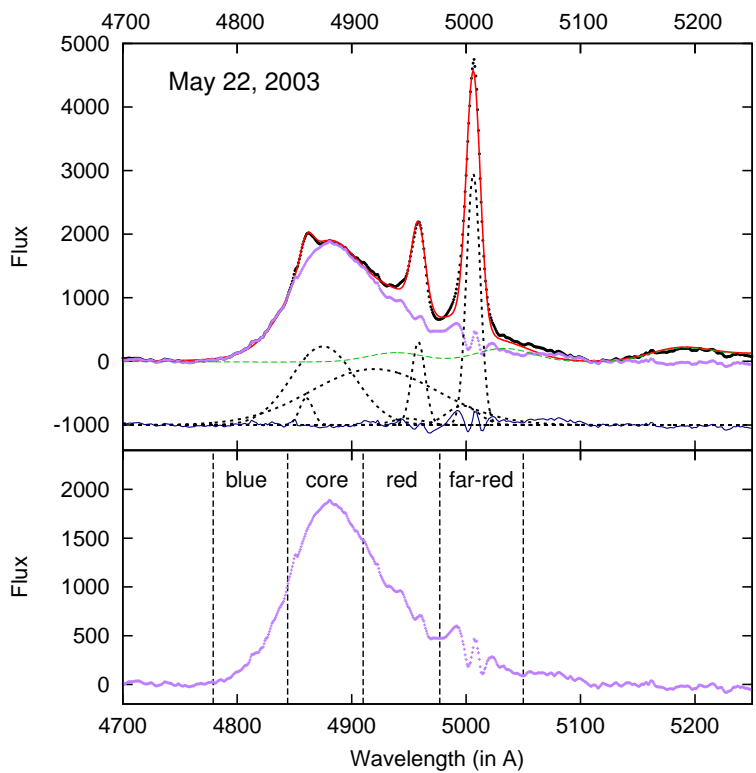}
\caption{{\it Upper:} Example of the best Gaussian fit of the continuum-subtracted spectrum around the H$\beta$ spectral
region (observed on May 22, 2003). Below the observed spectrum (dots), the best-fit (thick solid line), and the 
Fe II template (dashed line) are given. The residual (thin solid line), and the Gaussian broad and narrow 
components (dashed line) are shifted below for better visibility. {\it Bottom:} Broad H$\beta$ component, after
subtraction of narrow lines and Fe II lines. The segments (blue, core, red and far-red) of the line
are separated by vertical lines.} \label{gauss}
\end{figure}

Figure \ref{example} shows typical spectra of E1821+643  observed close to 
the minimum and maximum activity during the monitored period. It is obvious that in the spectra  of  
E1821+643 the broad emission lines of the hydrogen Balmer series and the big blue bump are most outstanding.  Also the narrow components of the Balmer lines and various forbidden lines,  typical for AGN (the most prominent [OIII]4959+5007 lines) are observed. As can be seen in Figure \ref{example}, there is no dramatic change
in the spectral energy distribution and Balmer continuum between the minimum and maximum stages.
Using the narrow line components we obtained the average redshift of $z = 0.2972\pm0.0002$ 
which we take as the rest frame of the host galaxy. We confirmed, as other authors found as well \citep[see][etc]{la08,ro10}, that the peaks of the broad Balmer line components are red shifted with respect to the narrow line components by $\sim$1000 km s$^{-1}$. 
The broad Balmer lines  have extremely asymmetric profiles, with red wings extending to Doppler velocities of at least $\sim$15000 km s$^{-1}$ relative to the rest frame wavelength (Table \ref{tab6}).

\subsubsection{Variability of the emission lines and the optical continuum}

\begin{figure*}
\centering
\includegraphics[width=16cm]{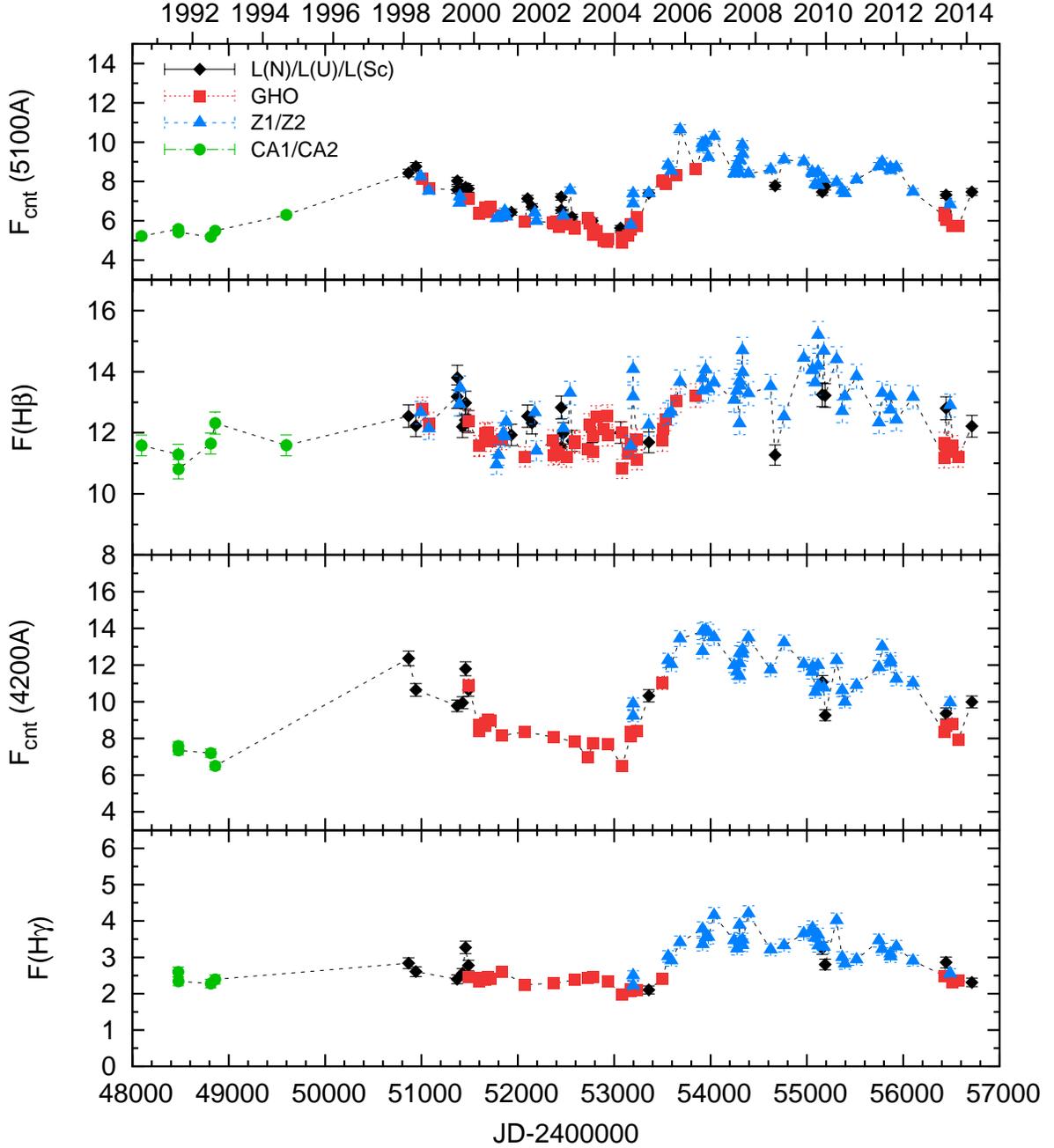}
\caption{Light curves (from top to bottom) for the continuum
 at 5100\AA, total H$\beta$, continuum at 4200\AA, and total H$\gamma$ line fluxes.
Observations with different telescopes are denoted with different
symbols given in the upper plot. The continuum flux is in units of
10$^{-15}$ erg cm$^{-2}$s$^{-1}$\AA$^{-1}$, and the line flux in units of
10$^{-13}$ erg cm$^{-2}$s$^{-1}$.} \label{lc}
\end{figure*}

\begin{figure*}
\centering
\includegraphics[width=6cm, angle=270]{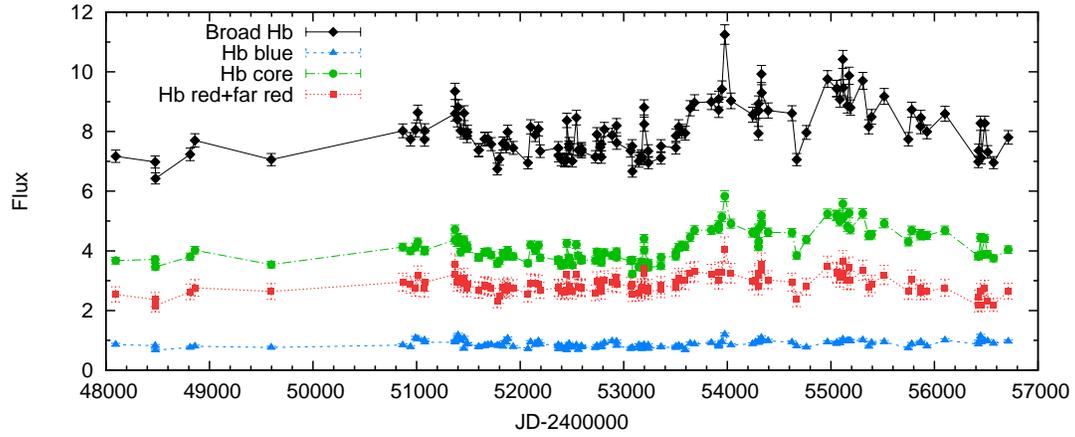}
\caption{Light curves of the broad-component and line-segment fluxes of H$\beta$. 
The fluxes are in units of 10$^{-13}$ erg cm$^{-2}$s$^{-1}$.} \label{lc-seg}
\end{figure*}

We analyzed  flux variations  in the continuum and lines using a total of 127 and 76 spectra covering the 
H$\beta$ and H$\gamma$ wavelength regions, respectively. In Table \ref{tab5} fluxes in 
continua at 5100 \AA \, and 4200 \AA, and  total H$\beta$ and H$\gamma$ lines are listed. Using these data we
plotted the light curves for the continuum at the rest wavelengths 5100 \AA \, and 4200 \AA, and for the H$\beta$ and H$\gamma$ lines (see Figure \ref{lc}).  It is obvious that the fluxes in the continua and lines are varying during this monitored period (1990--2014).  In the continuum light curves, and with some possible delay
in the H$\beta$ and H$\gamma$ light curves, the same flare-like events (1-3 from Table \ref{flare}) as in the R-band photometric light curve (see Figure \ref{fig1}) are seen.
In Figure \ref{lc-seg} the light curves for the broad H$\beta$ line component and its different line-segments are
given. Both fluxes in  the broad H$\beta$ line component and in its line-segments (blue, core, red+far red, 
Table \ref{segm}) vary quasi-simultaneously and local maxima, approximately close to the flare-like events 1-3, are also traced. In Figure \ref{lc-seg}, there are some indications of the presence of the shorter time-scale fluctuations (flares) in the light curves of the total-line and line-segment fluxes. These flare-like features become clearer in the artificially generated light curves (see Section 4.1). One can suspect that a fraction of short-time flares arise from the uncertainties in the relative flux calibration (i.e. unification of the spectral data, see Sections 2.4 and 2.5), however, the short-time flares can be also detected in spectrophotometic curves as a consequence of short time variability of the nucleus. For example, in several other objects, 
the short-time flares have been registered in line/continuum light curves \citep[e.g. Ark 564 and Arp 102B, see][]{sh12,sh13}. Since in case of this object, the short-time flares are not within the frame of the error-bars, thus it seems that from time to time the short-time flare events are present in the light curves of E1821+643.

For the variability estimates of the line and continuum fluxes,
we used the method given by \cite{ob98}.  In Table \ref{var} 
we give parameters that describe the variability of the continuum and
total line fluxes, i.e. N is the
number of spectra, $F$ denotes the mean flux over the whole
observing period, and $\sigma(F)$ is the standard deviation, and
$R$(max/min) is the ratio of the maximal to minimal flux in the
monitoring period. The parameter $F$(var) is an inferred
(uncertainty-corrected) estimate of the variation amplitude with
respect to the mean flux, defined as:
$$ F({\rm var})= [\sqrt{\sigma(F)^2 -e^2}]/F({\rm mean}) $$
$e^2$ being the mean square value of the individual measurement
uncertainties for N observations, i.e. $e^2=\frac{1}{N}\sum_i^N
e(i)^2$ \citep{ob98}.  From Table \ref{var} it follows that the fluxes in the continuum (at rest 5100 \AA \, 
and 4200 \AA) and total  H$\gamma$ line changed for about 2 times, while in the  H$\beta$ flux only for $\sim$1.4 times. 
The difference in the line flux variations between  H$\beta$ and H$\gamma$ may be caused by different dimensions of 
emitting region for these two lines (i.e. H$\gamma$ emitting region is significantly smaller than H$\beta$ one). In addition,
the smaller variability in the H$\beta$ flux may be caused by the contribution of the constant fluxes from narrow  
lines [OIII]4949,5007.  The amplitude of variability F(var) is $\sim$19\% for the continuum and total 
H$\gamma$ line, and $\sim$7\%  for the total H$\beta$ line. 
The fluxes of the broad H$\beta$ segments (blue, red, 
far-red wings, and core) vary quasi-simultaneously with F(var)$\sim$11-12\% (Table \ref{var}, Fig. \ref{lc-seg}).

\subsubsection{The lines and continuum flux correlations}

\begin{figure}
\centering
\includegraphics[width=11cm]{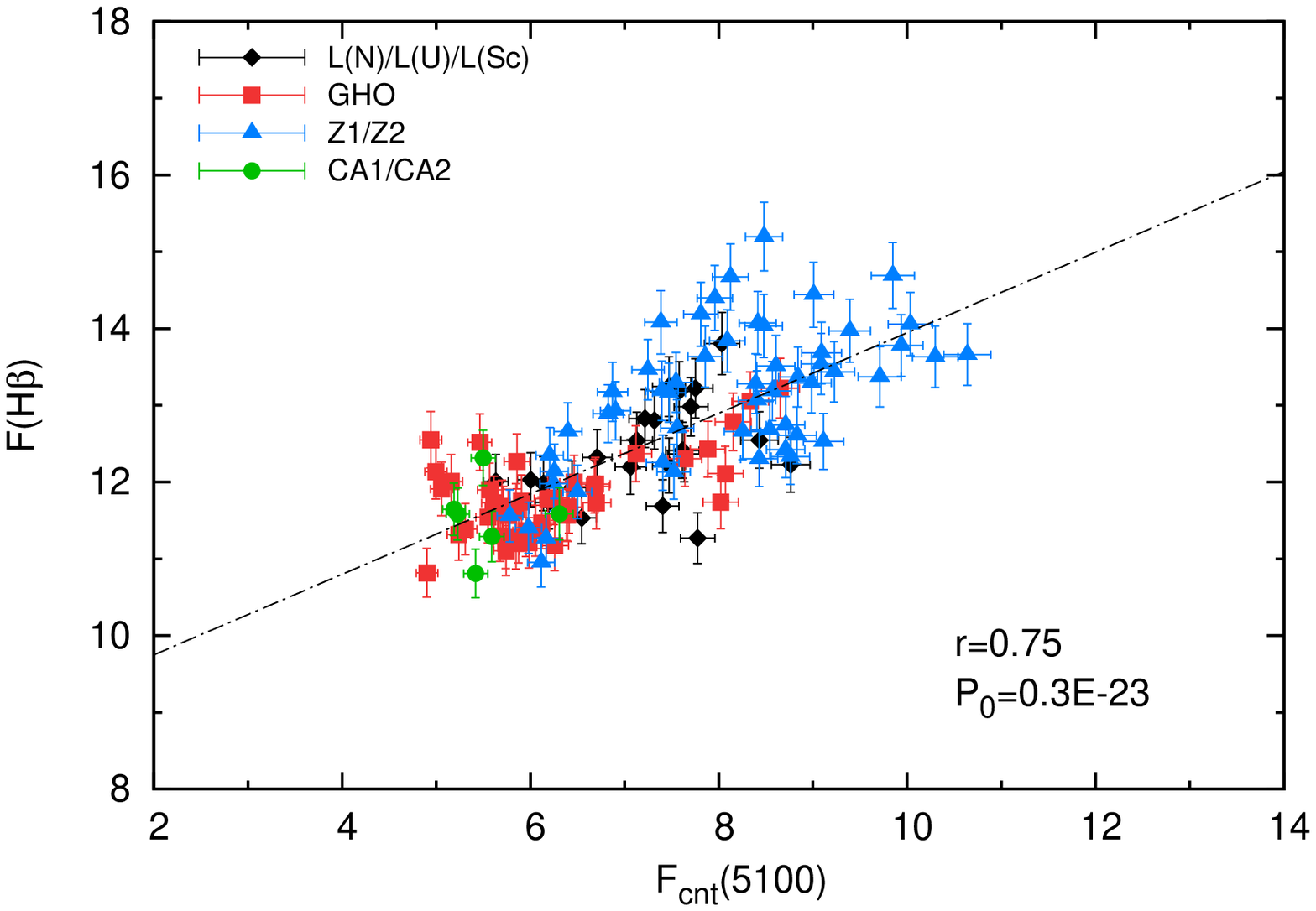}
\includegraphics[width=11cm]{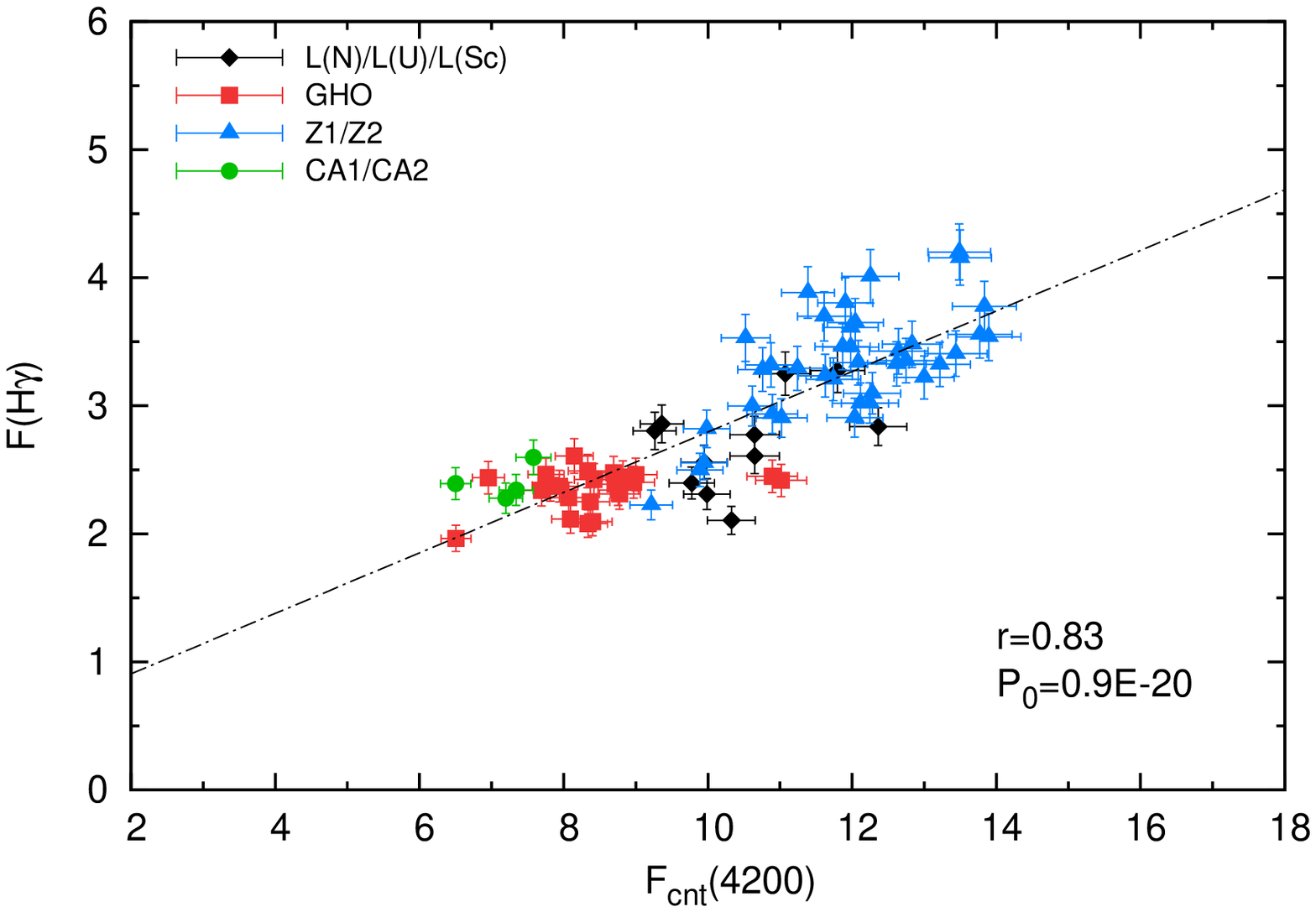}
\caption{H$\beta$ vs continuum at 5100\AA \, flux (upper) and H$\gamma$ vs. continuum 
at 4200\AA\ flux (bottom). The continuum flux is in units of
$10^{-15} \rm erg \ cm^{-2} s^{-1} \AA^{-1}$, and the line flux in units of
$10^{-13} \rm erg \ cm^{-2}s^{-1}$. Observations with different
telescopes are denoted with different symbols given in the upper left.
The correlation coefficients and the
corresponding p-values are also given.} \label{Hb_cnt}
\end{figure}

\begin{figure}
\centering
\includegraphics[width=11cm]{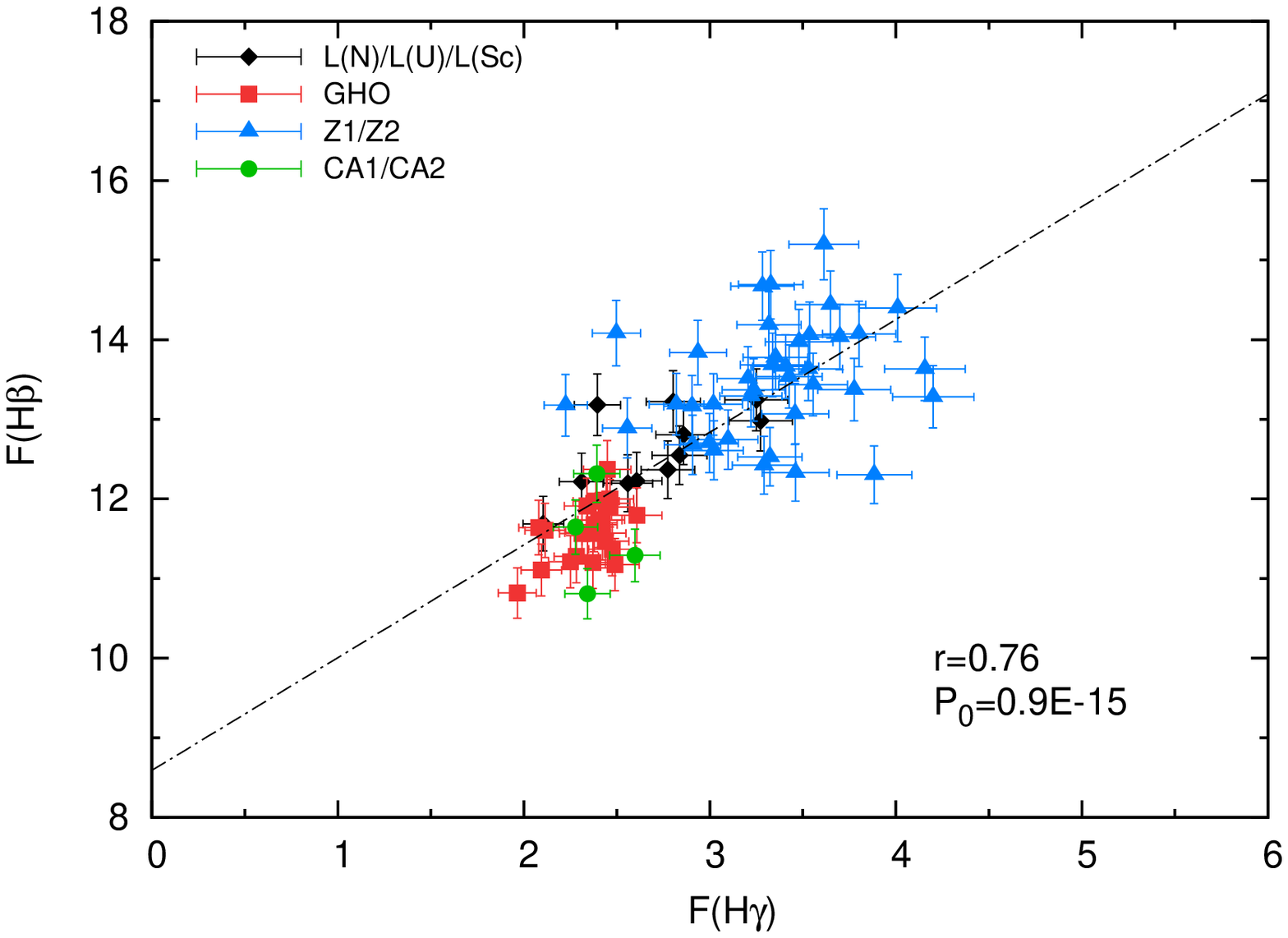}
\includegraphics[width=11cm]{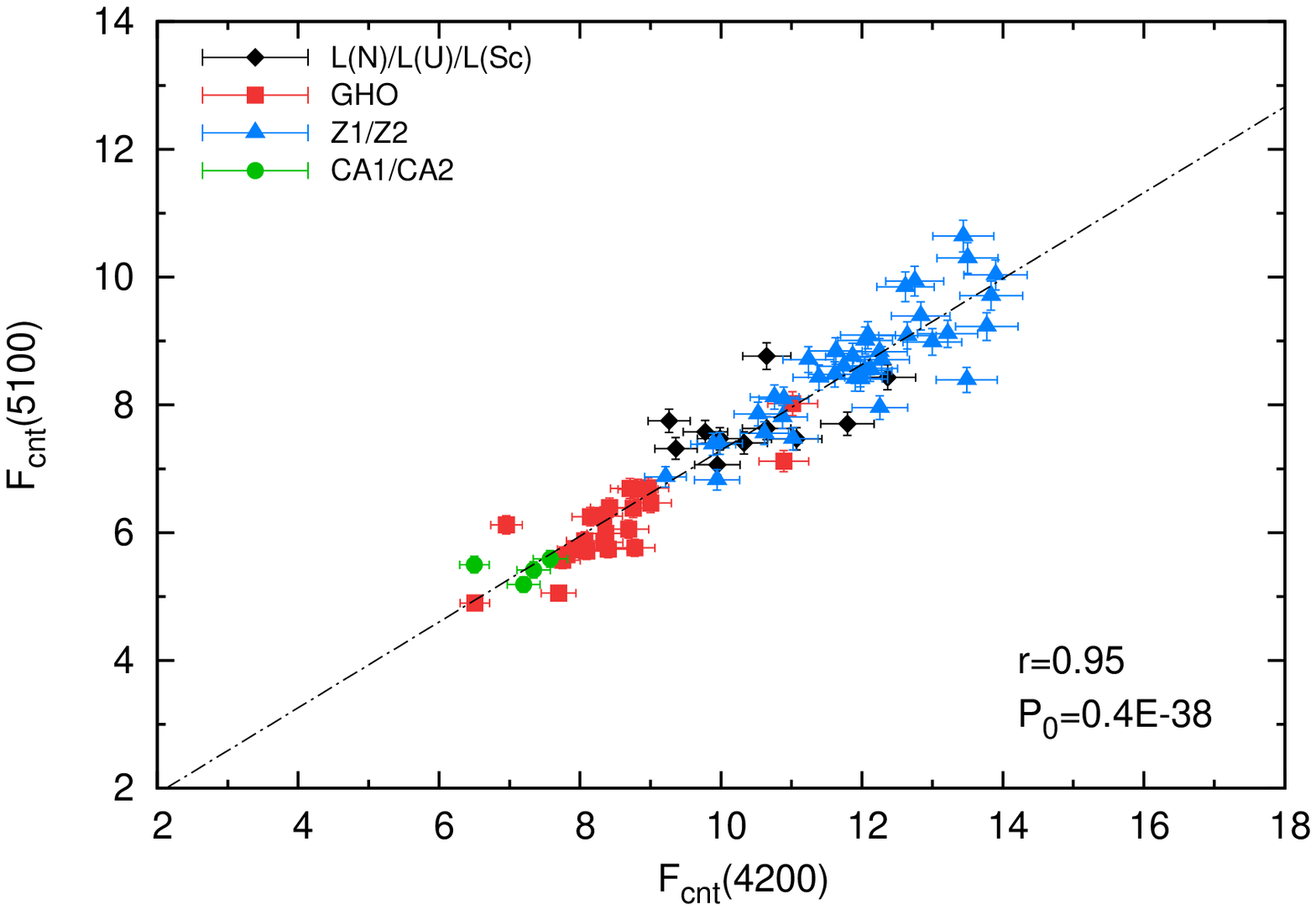}
\caption{The H$\beta$ vs. H$\gamma$ flux (upper) and continua 5100\AA \ vs. 4200\AA\ flux (bottom). The continuum flux 
is in units of $10^{-15} \rm erg \ cm^{-2} s^{-1} \AA^{-1}$, and the line flux in units of
$10^{-13} \rm erg \ cm^{-2}s^{-1}$. Observations with different
telescopes are denoted with different symbols given in the upper left.
The correlation coefficients and the corresponding p-values are also given.} \label{fig9}
\end{figure}

\begin{figure}
\centering
\includegraphics[width=11cm]{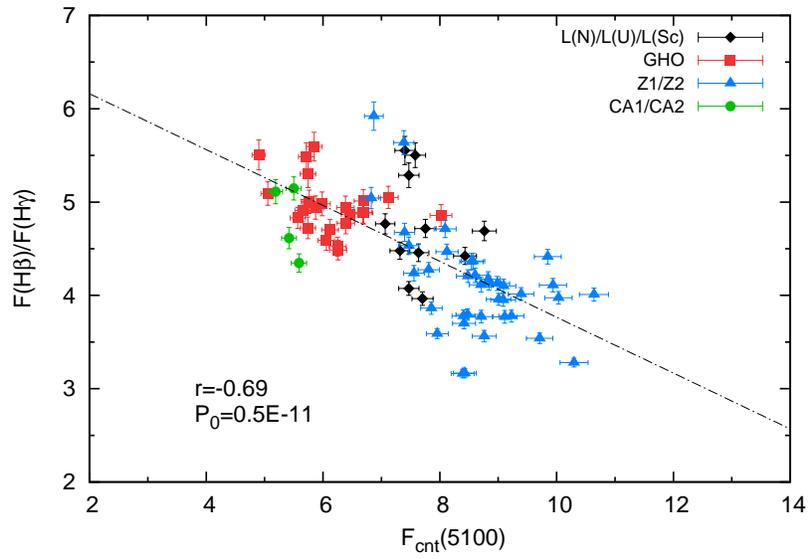}
\caption{H$\beta$/H$\gamma$ flux ratio vs. continuum flux at 5100\AA . The continuum flux is in units of
$10^{-15} \rm erg \ cm^{-2} s^{-1} \AA^{-1}$. Observations with different
telescopes are denoted with different symbols given in the upper right.
The correlation coefficients and the
corresponding p-values are also given.} \label{fig10}
\end{figure}

\begin{figure}
\centering
\includegraphics[width=11cm]{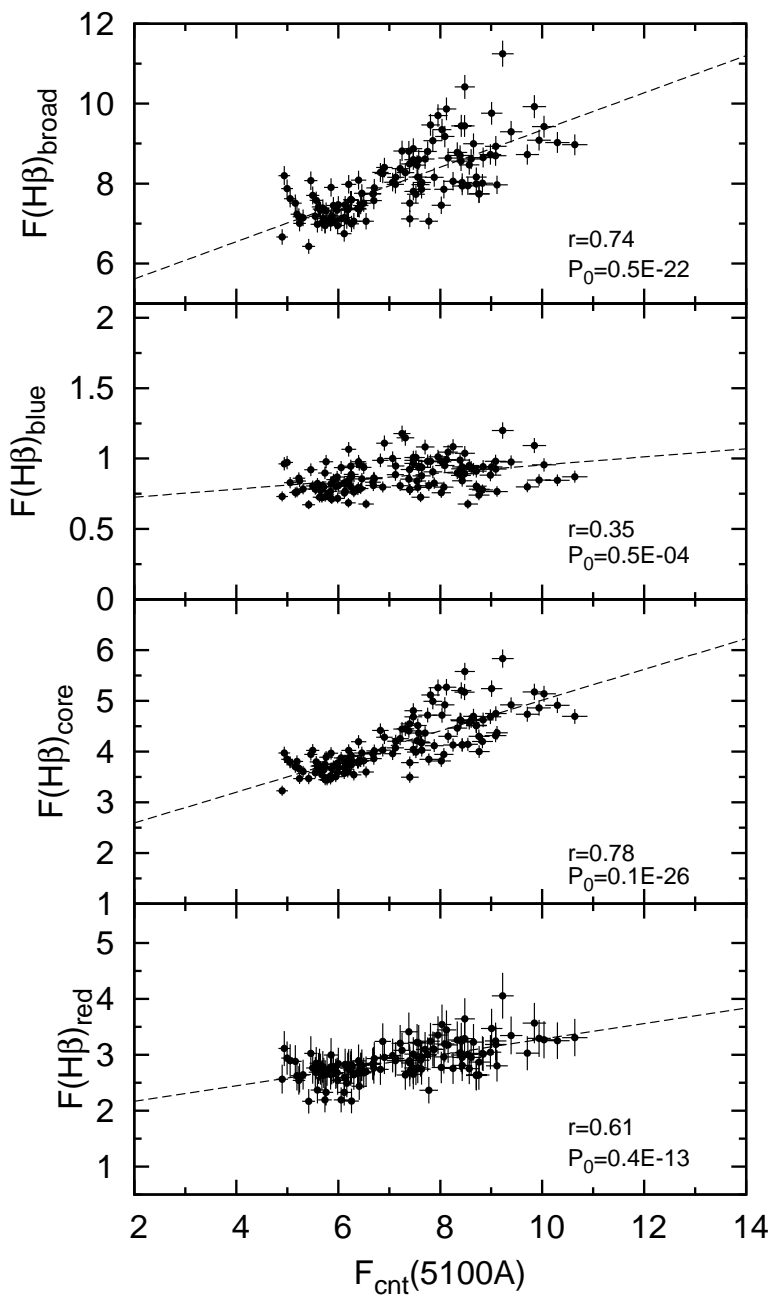}
\caption{The broad H$\beta$ component and line-segment fluxes (blue,
core and red) vs. continuum flux at  5100 \AA , respectively from top to bottom. The
continuum flux is in units of $10^{-15} \rm erg \ cm^{-2} s^{-1}
\AA^{-1}$, and the broad H$\beta$ component and line-segment fluxes are in units of $10^{-13} \rm erg \
cm^{-2}s^{-1}$. The correlation coefficients and the corresponding
p-values are given in the bottom right corner.}
\label{wings_c}
\end{figure}

\begin{figure}
\centering
\includegraphics[width=11cm]{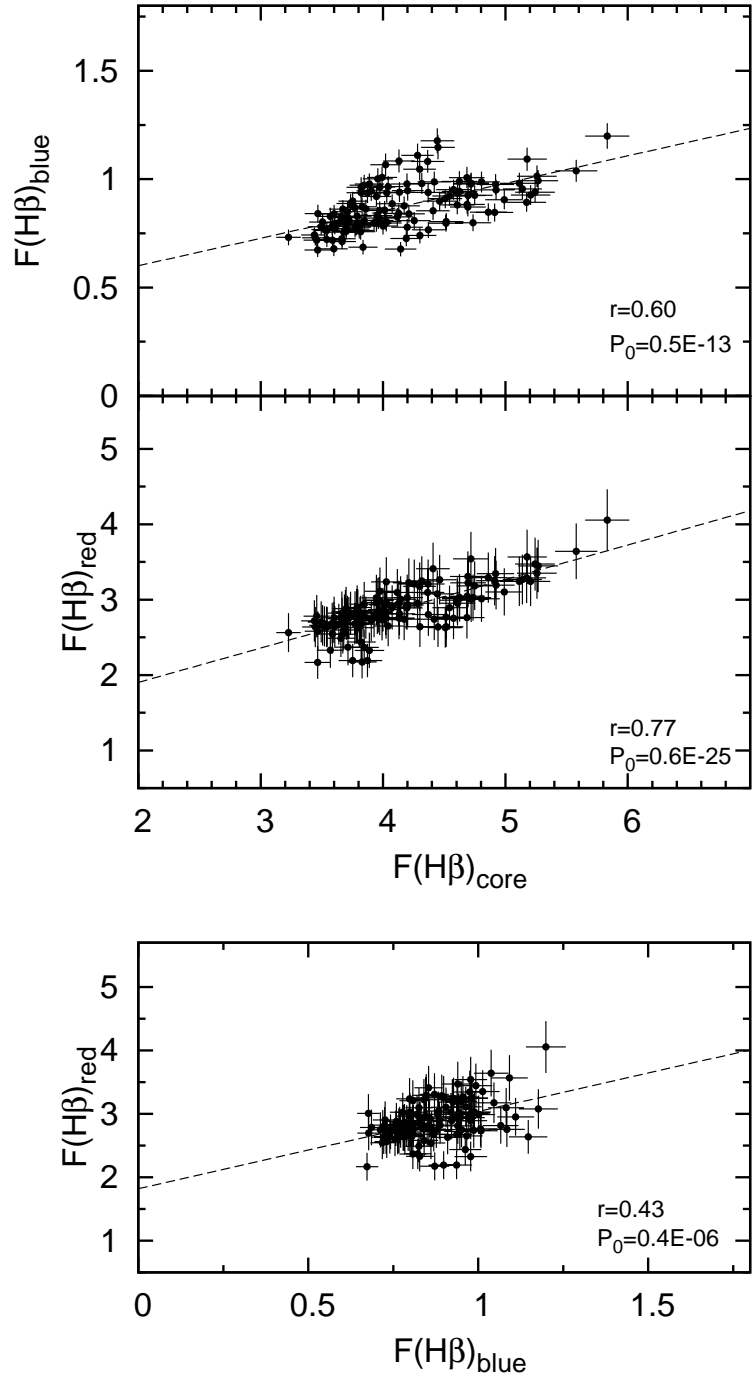}
\caption{H$\beta$ line-wing fluxes (blue, red) vs.
line-core flux (two upper panels), and red wing vs. blue wing flux (bottom panel).
The line-segment fluxes are in units of $10^{-13} \rm erg \
cm^{-2}s^{-1}$. The correlation coefficients and the corresponding
p-values are given in the bottom right corner.}
\label{wings}
\end{figure}

Figure \ref{Hb_cnt} shows the relationship between the total line (H$\beta$, H$\gamma$) and the adjacent continuum fluxes (at 5100 \AA \, for H$\beta$ and 4200 \AA \, for H$\gamma$).  The correlation coefficients in both cases are quite high ($\sim$0.8), indicating that the ionizing continuum is a good extrapolation of the optical continuum. We should note here that
there is a slightly better correlation between the flux of H$\gamma$ and the continuum at 4200 \AA \, 
than between H$\beta$ and  continuum flux at 5100 \AA . This indicates a faster response of the
H$\gamma$ line flux to the corresponding continuum changes, than seen in H$\beta$.
In Figure \ref{fig9} the relationship between H$\beta$ and H$\gamma$ fluxes (upper panel)
and between the  continua   at 4200 \AA \, and 5100 \AA\ (bottom panel) are given.
As can be seen, there are good  correlations 
between the ratio of the H$\beta$ and H$\gamma$ total line fluxes (r$\sim$0.76), and even better
between two continua (r=0.95). 
Figure \ref{fig10}  gives the correlation between the total line-flux ratio H$\beta$/H$\gamma$ vs. continuum 
flux at 5100 \AA . A strong anti-correlation (r$\sim$0.7)
is observed in this case. This also indicates
faster and stronger reaction of the H$\gamma$ line flux to the continuum than the H$\beta$ one.
Finally, in Figures \ref{wings_c} and \ref{wings} the correlations between the broad H$\beta$ line, its 
line-segments and continuum flux, as well as correlations between different broad line H$\beta$ line-segments 
relative to each others are given.  There are significant correlations (r=0.6-0.8) between the line core and 
red wing flux with the continuum flux, however the correlation of the blue wing and continuum fluxes is smaller (r=0.35) and less statistically significant.

\subsubsection{Time-lag analysis}

There are several classes of methods for handling the problem of
irregular data sets time-lag analysis. Perhaps the oldest class of
estimators is standard interpolation (using linear and cubic-spline) of
observations in order to create time series on a regularly spaced grid.
This method leads to significant
reduction in variance toward the high-frequency range of the estimated
power spectrum. When there are interests in phenomena on  smaller
timescales  (relative  to the mean sampling interval) such effects
should be taken into account. Beside this, the persistence (memory) of
irregularly time series  is strongly overestimated when using the
standard  (linear and cubic spline) interpolation  approach \citep[][]{re11}.

Our data sets are irregular, thus for the time-lag analysis of the
spectroscopic and  photometric light curves we applied three methods: 
(i) the Discrete Correlation Function \citep[DCF;][]{EK188}, (ii) the z-transformed Discrete Correlation Function \citep[ZDCF;][]{A97,A13}, both methods from the class of slotting time-lag estimators, and (iii) the Stochastic Process Estimation for AGN Reverberation (SPEAR) by \citet{zu11}, recently developed method that is a model-based estimator that uses a damped random walk model \citep[DRW model][]{k09,koz10,ma10}.
For more details on all three methods and their differences and advantages see a recent paper 
by \citet[][and references therein]{ko14}.

We calculated the DCF and ZDCF  functions over the major time range of our monitoring 
campaign (the points before 1998 are excluded due to extremely poor sampling). 
The ZDCF lags are taken to be the peaks of the CCFs, while the uncertainties are addressed
using the Monte Carlo method \citep{A97, A13}.
The DCF lags, DCF coefficients and their errors are calculated using 
the MATLAB bootstrap toolbox for time series analysis. For this, we constructed 1000 clones 
of our  light curves. Each clone is made by choosing random samples with 
replacement from mother curve, i.e. each observation is selected separately at random from the original
dataset. The number of elements in each bootstrap clone equals the
number of elements in the original data set. Then the centroid of DCF was computed
for each pair of clones so we can construct bootstrap vectorial
statistics  of time lags and centroid of DCF. The mean values are
chosen as the final DCF lag and coefficients results, and the
uncertainties are calculated as standard deviations of their bootstrap
distributions.

\begin{figure}
\centering 
\includegraphics[width=17cm]{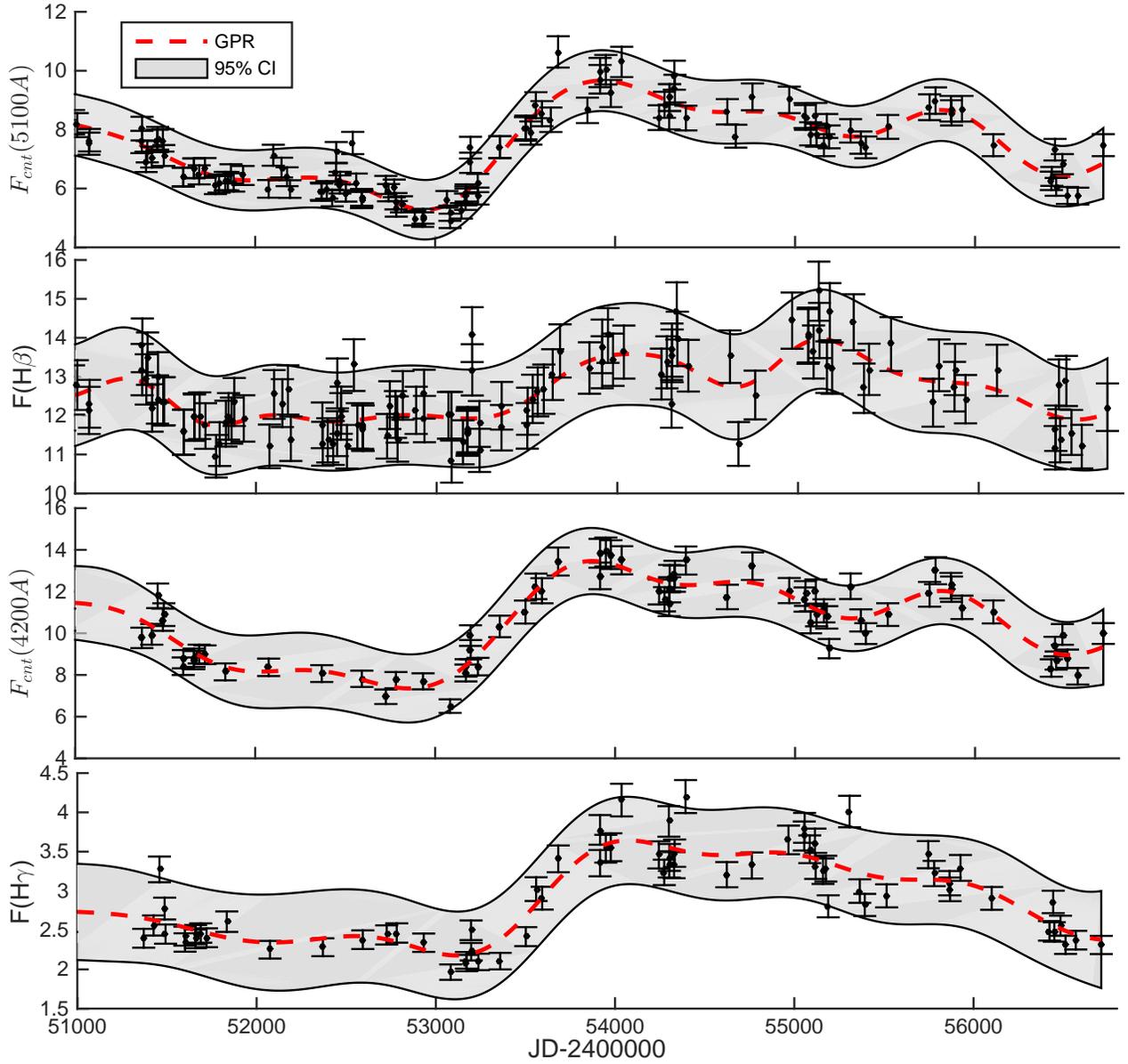}
\caption{Comparison between the GPR generated (dashed line) and observed light curves 
(circles with errorbars) of the continuum at 5100\AA, H$\beta$, the continuum at 4200\AA, and H$\gamma$ (from top to bottom). 
The shaded band represents $95\%$ confidence interval (CI) for the GPR predicted curve.} \label{gpr} 
\end{figure}

 Since our light curves have large mean sampling, we attempt to
 probe time-lags of artificial time series with 
 better sampling and which points are "predictions" obtained from original time series.
 For this task we  employed Gaussian process regression (GPR) for noisy data \citep{EK88}.
 GPR generates data such that if we observe their values, they would
 follow a multivariate Gaussian distribution. The main constituents 
 of a GPR are its covariance and mean functions.
With covariance function  we encode correlations (relations or
similarities) between different data points in the process.
We specified a GPR  for our artificial time series as follows: a constant
    mean function, with initial parameter set to mean of original time
    series, and an isotropic squared exponential covariance (kernel)
    function. The covariance function takes two hyperparameters, 
    a characteristic length-scale and the standard deviation of the original signal. 
    The length-scale was set to 150.  Fig. \ref{gpr} gives the comparison between the 
    GPR generated and observed light curves of the continuum at 5100\AA, H$\beta$, 
    the continuum at 4200\AA, and H$\gamma$ (from top to bottom).

\begin{figure} 
\centering
\includegraphics[width=5cm]{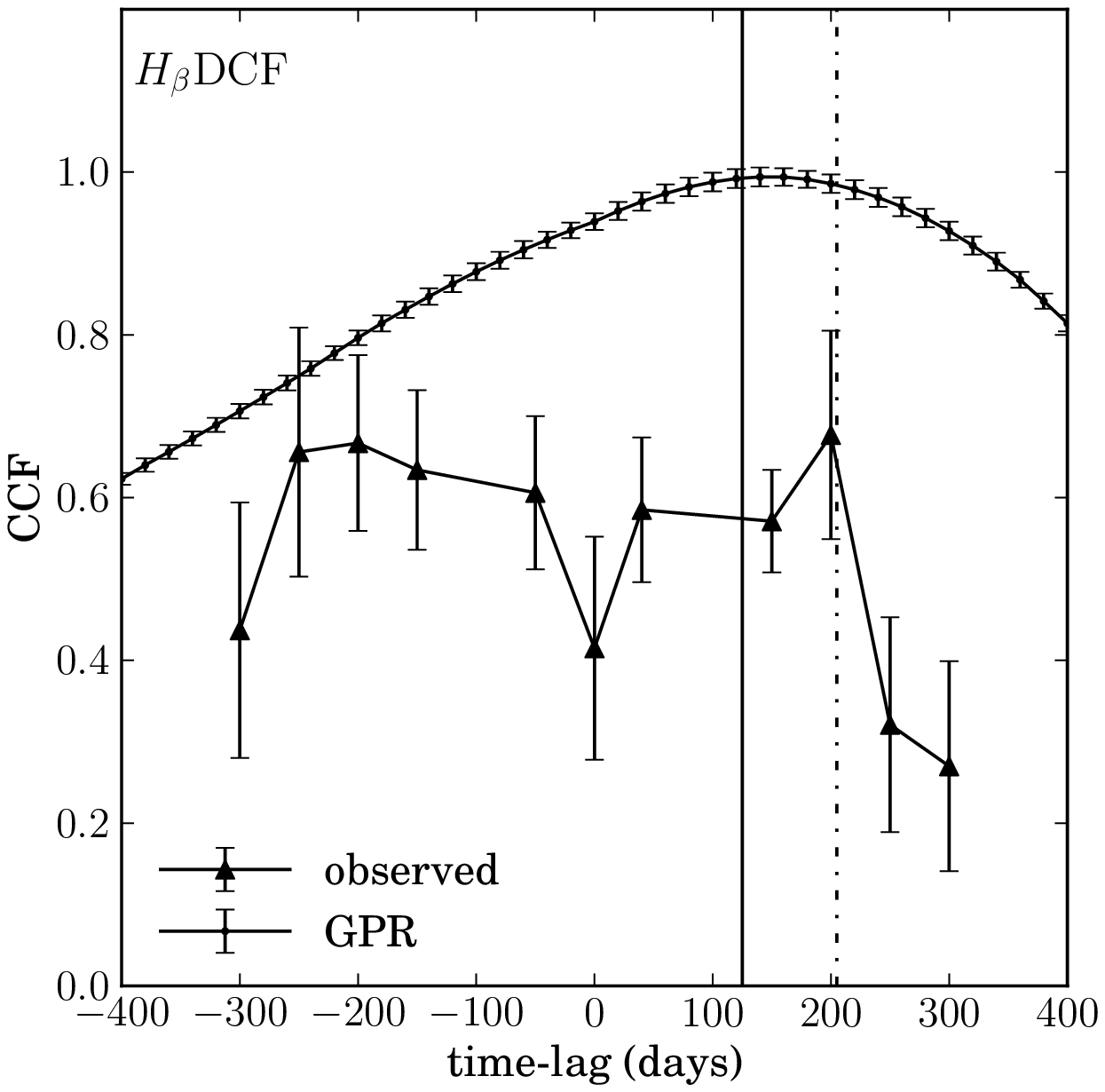}
\includegraphics[width=5cm]{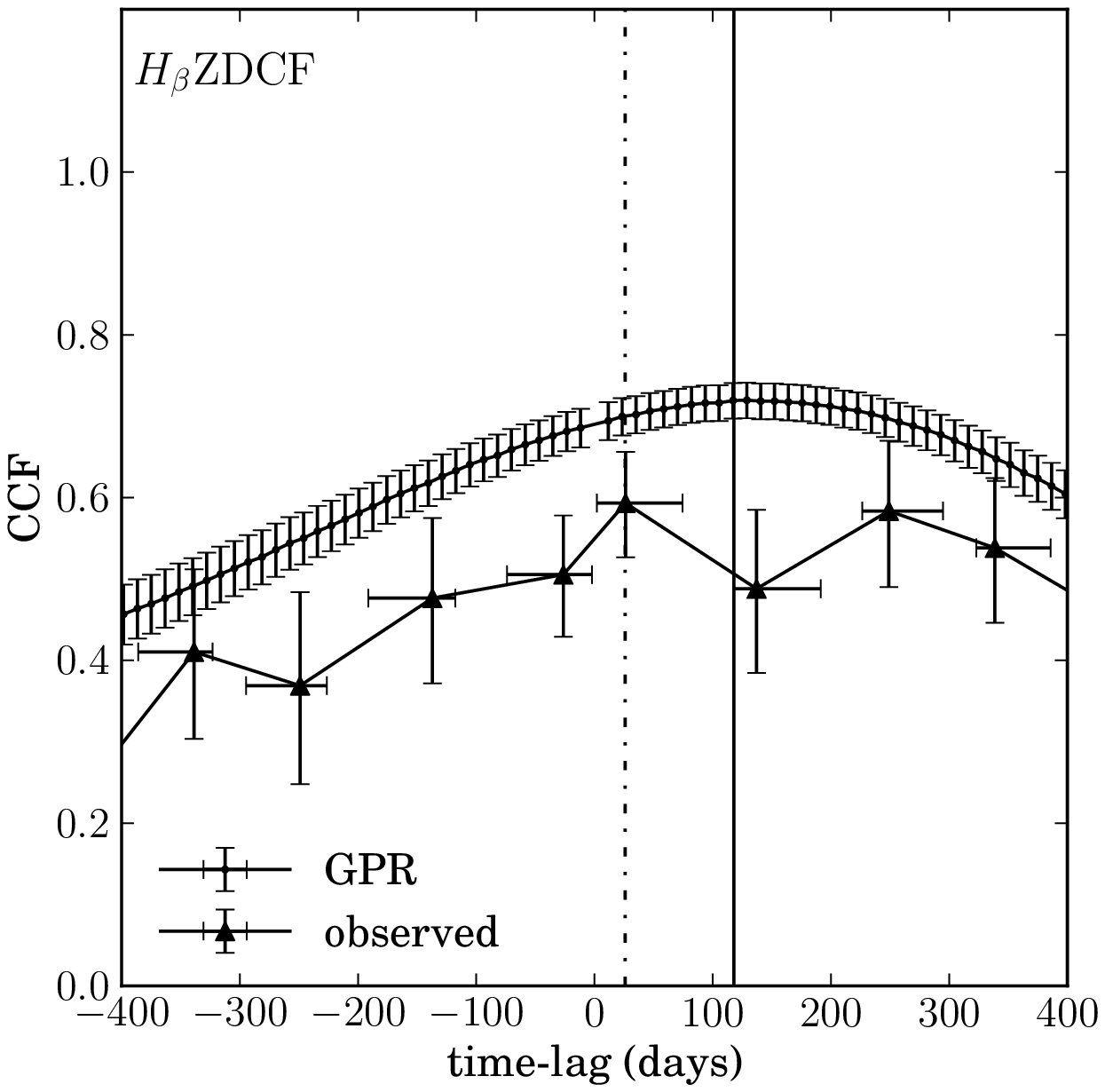}
\includegraphics[width=5cm]{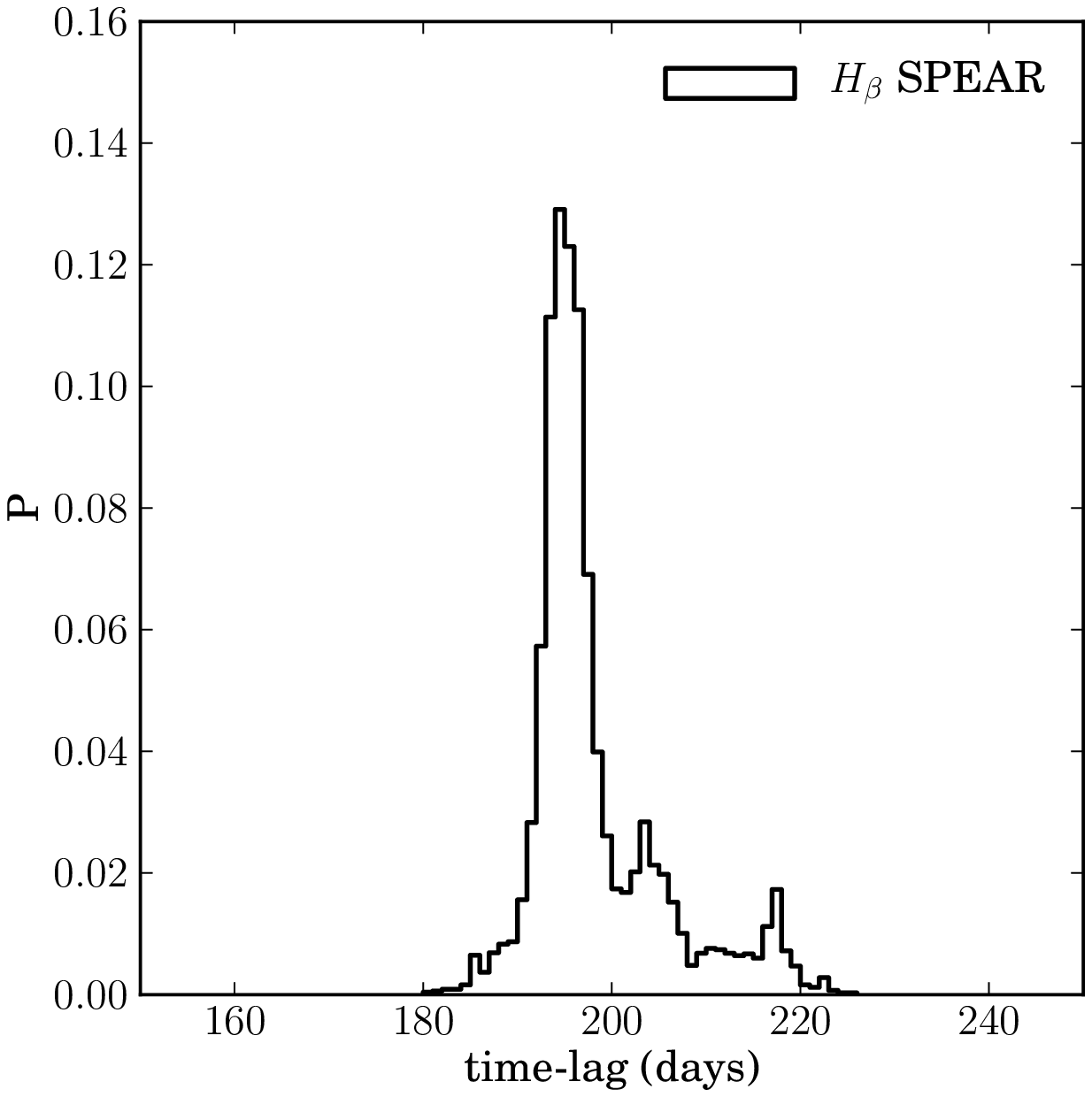}
\includegraphics[width=5cm]{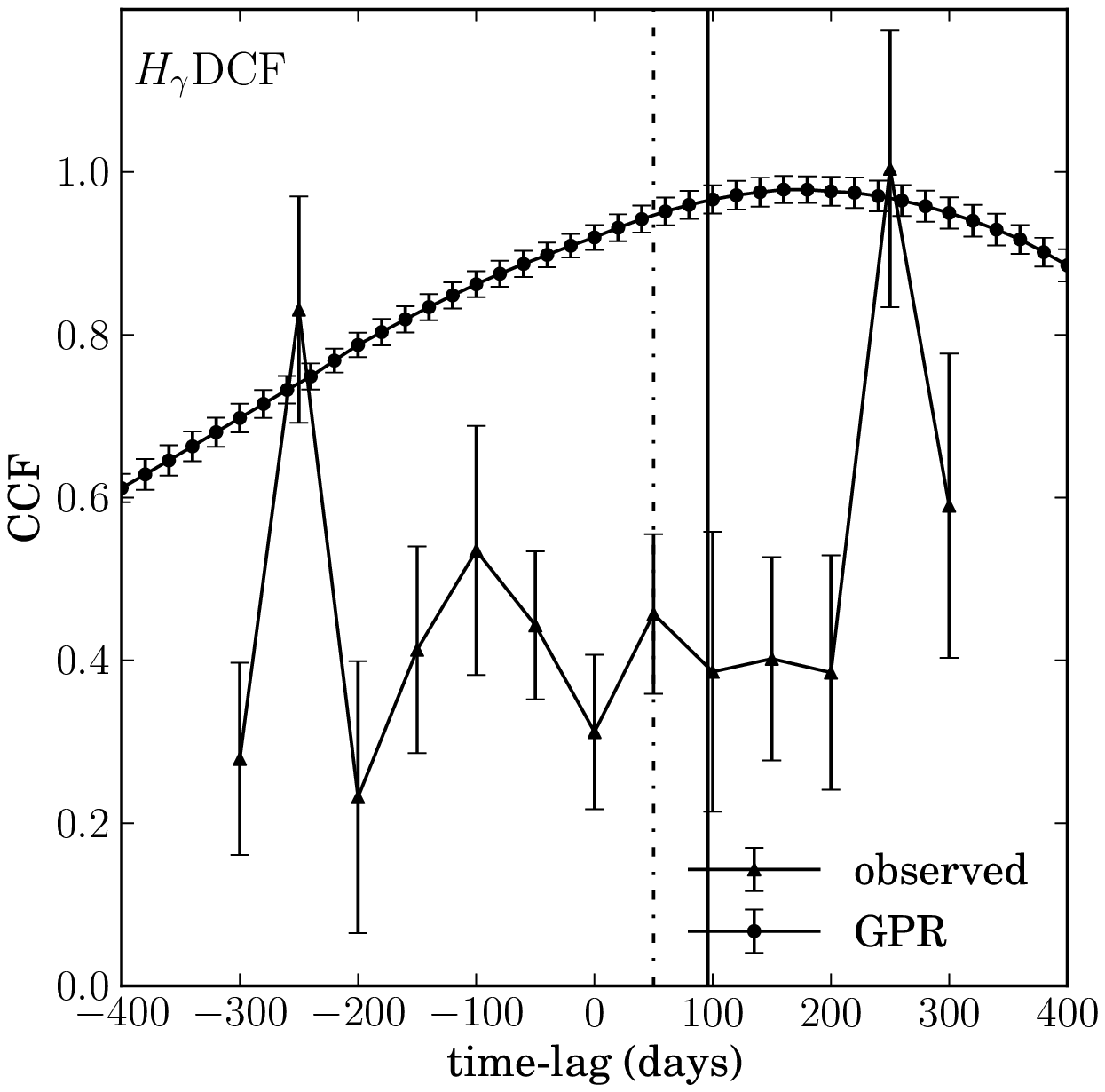}
\includegraphics[width=5cm]{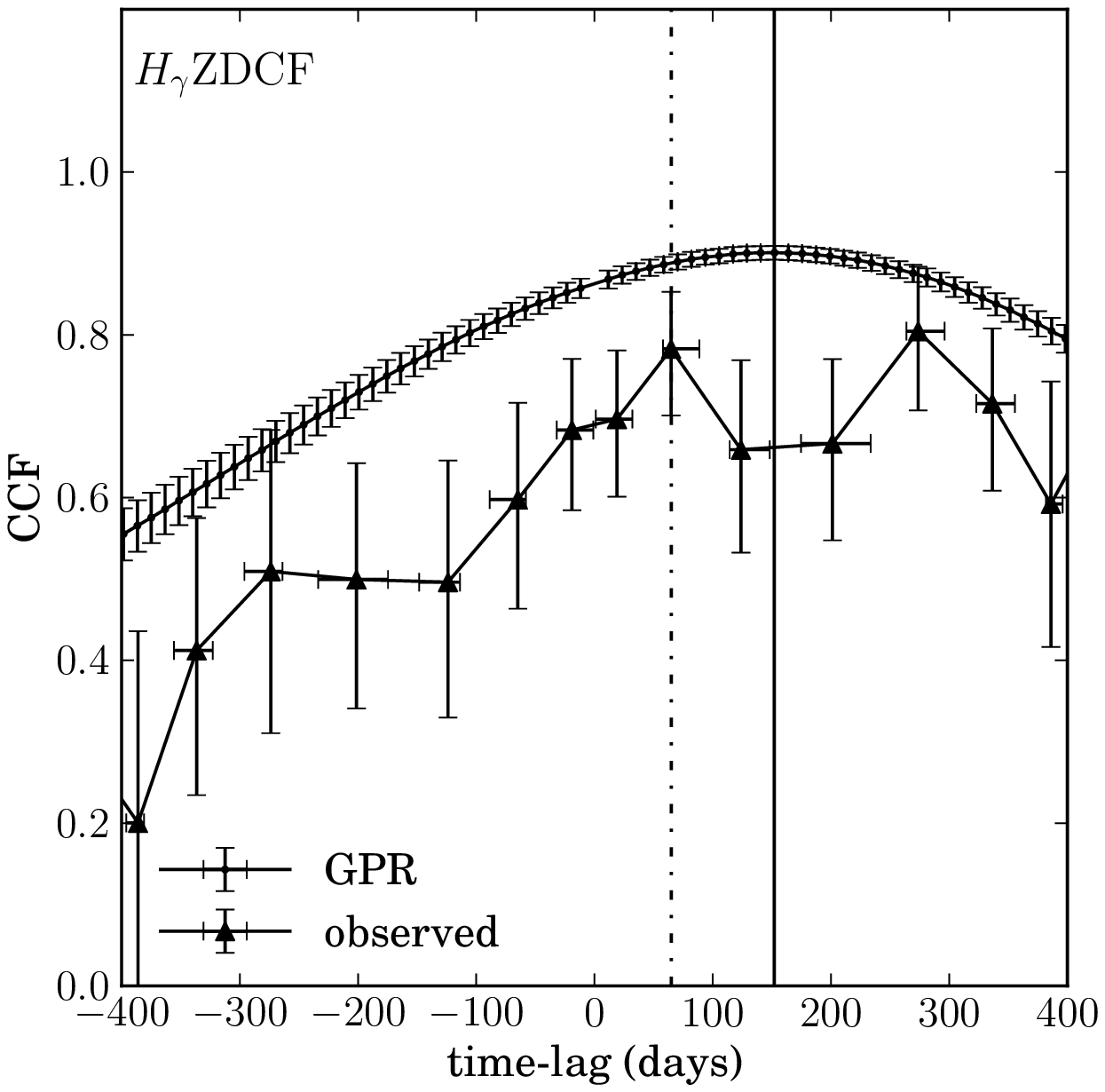}
\includegraphics[width=5cm]{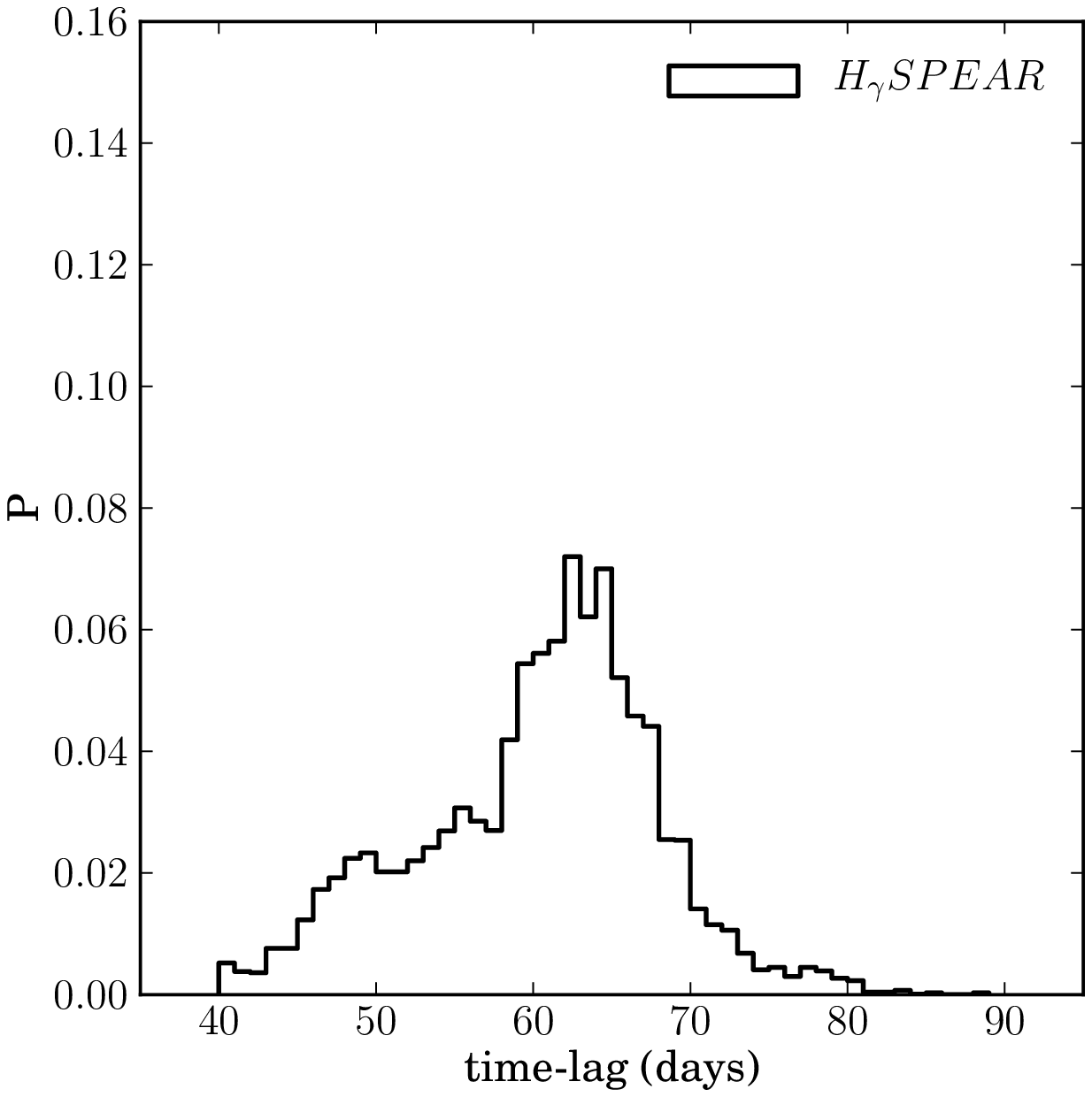}
\caption{The derived CCFs of the observed (triangles) H$\beta$ (upper panels) and H$\gamma$ (bottom panels) time-lags from the DCF (left) and ZDCF (middle) methods, and the 
equivalent probability distribution P from the SPEAR method (right). The solid continuos curves
give for comparison the derived CCFs of the corresponding  GPR generated artificial light curve. The vertical lines denote the obtained time-lag in case of the observed (dash-dotted) and GPR generated (solid) data.} \label{fig-ccf} 
\end{figure}

Table \ref{ccf} summarizes the results of all three time-lag estimators we
employed on both observed and GPR generated artificial light curves of H$\beta$ and H$\gamma$
and the corresponding continuum light curves. The GPR artificial light curves have uniform  
and better sampling. Fig. \ref{fig-ccf} gives the derived CCFs (triangles) of H$\beta$ (upper panels) and H$\gamma$ (bottom panels) time-lags from the DCF (left panels) and ZDCF (middle panels) methods, and the equivalent probability distribution from the SPEAR method (right panels). The derived CCFs of the corresponding  GPR generated artificial light curve are also given for comparison. The large number of points in the GPR
light curves leads to better constrained uncertainties in the case of the DCF and ZDCF methods, due to consequently larger number of binned points. 

In case of the DCF and ZDCF methods, the derived CCFs
obtained from the observed light curves (left and middle panels in Fig. \ref{fig-ccf}) typically 
contain two peaks of similar amplitude that (considering the error bars) are not statistically 
distinguishable. In Table \ref{ccf} we listed the values of the first clearly visible peak. With this we demonstrated that the reliable lags cannot be recovered directly from this data sets using the DCF and ZDCF methods. Another point that can be seen from Table \ref{ccf} is that the ZDCF gives for both
H$\beta$ and H$\gamma$ the time-lags that are similar to the median sampling interval, 
and thus might be considered as spurious. However, this may also be a
coincidence as other authors have obtained time-lags similar to the median sampling interval
\citep[see e.g.][]{de09}.

Also we should note here, that the obtained time-lags can be influenced by the 
relative flux calibration (i.e. unification of the spectral data, see Sections 2.4 and 2.5) 
of the spectra and the short-time flares, which can be seen in the 
light curves of the continua and broad lines.  Data calibration procedure in 
constructing the light curves can affect the traditional CCFs analysis, and can 
have greater effect on the lag analysis of the more sparsely sampled H$\gamma$ -- continuum light curves. 
That is why we additionally applied the SPEAR method, which has been tested
for different calibration effects, and it has been shown that the impact on the estimated lag is generally negligible \citep[see][]{zu11}. Additionally, as can be seen in Fig. \ref{gpr}, the GPR procedure allows us to avoid the influence of the flare-like local peaks. 

The SPEAR method treats large time-gaps in light curves in a statistically based approach, 
and considers the impact on the uncertainties of the time-lag. Our data sets have large 
time-gaps in the light curves and flare-like peaks, thus the obtained ZDCF and DCF curves 
are degenerated, while the SPEAR time-lag probability distributions are well defined 
with no deterioration. Therefore the obtained SPEAR lags are preferred. 

As a summary of our CCF analysis, we can  state that there is a large difference between the H$\beta$ and H$\gamma$ lags obtained from the SPEAR method (in both cases, for the observed and GPR time-series). 
It seems that the H$\gamma$ emitting region is smaller than the H$\beta$ one, that is also in agreement with dependence of the flux ratio of lines as a function of the continuum (see Figure \ref{fig10}). Further in the text, we will use the values of lags obtained from the SPEAR method applied on the GPR generated  artificial light curves, i.e. the lag for H$\beta$ $\sim120$ and for H$\gamma$ $\sim60$ days.

\subsubsection{Periodicity}

As noted in sections 3.1 and 3.2.1, 3-4 maxima are visible in the photometric and spectral light  curves 
(see Figure \ref{fig1}, the maxima are denoted with vertical ticks). 
This motivated us to search for periodicities in different  light curves. 

We applied periodogram analysis to test whether our time series contain only noise or some
periodic components are present. In the case of presence of semi-periodic or  even non-periodic components,
the periodogram can show more than one prominent peak. Another problem is that the mean
values are not always good estimator of the mean of periodogram underlying function, causing
problems as aliasing.
In order to avoid these problems, we applied several techniques to our time series: (i) "Generalized" Lomb-Scargle periodogram\citep[GLS][]{l76,sc82},
(ii) multiterm periodogram routine  \citep[MP][]{va12,iv14}, and (iii) Bayesian Generalized Lomb-Scargle periodogram \citep[BGLS][]{mo15}.
GLS treats the problem of the mean overestimation  by adding constant offset term to the model. MP treats the problem if the data have
hidden variability that is more complex than single sinusoidal. BGLS includes both weights and a constant
offset in the data.

Figure \ref{period1} compares GLS of our time series. The light curves of continua at 4200 \AA\ and
5100 \AA, and the H$\beta$ and H$\gamma$ lines have a strong peak corresponding to the period of about 
4500 days, that is the only statistically significant peak. Periodograms
of both H$\beta$ and H$\gamma$ lines show another two peaks 
around 1300 days and  780 days ($\omega\in[0.003,0.005]$), but they are bellow
the 1\% and 5\% significance levels. The peak of about 780 days ($\sim$ 2 years) can be a consequence of
observation conditions. MP and BGLS techniques give a peak that is also about 4500 days for
both continuum and both line light curves.
Periodograms of fluxes of photometric curves show 3 prominent and statistically significant peaks
around 4000, 1850, and 1200 days (Figure \ref{period1}). 

Comparing the obtained periods with observed peaks in light 
curve, one can  see that the 4000 (photometric) and 4500 (spectroscopic)
day periods are  representing only one cycle period and it is hard to conclude that we found a good evidence for periodic behavior with approx. 4000 day period. However, the 1200 day period, that is
associated with the 4 flare-like events in the photometric light curve represents an evidence
for a periodic variability in the photometric light curve, as well as in the spectroscopic light curves (1300 days, even it has a small significance level).  The period of 1850 days found in the photometric light curve, is not present in the spectroscopic light curves. With current analysis of the variability in the continuum and line fluxes we cannot find any associated physical phenomena with this period.

The mean  sampling period of the continuum and light curves is about 50 days, so it is possible that any shorter period is hidden in the poor data sampling since MP and BGLS could not retrieve them with larger significance.

\subsubsection{Line profile variation}

Using the continuum and narrow line subtracted spectra, we constructed the mean and root-mean-square (rms) 
line profiles of the broad H$\beta$ and H$\gamma$ components (see Figure \ref{mean}), which shows that the broad lines  have negligible changes in their profiles, i.e. during a long period of 24 years the H$\gamma$ and H$\beta$ line profiles have a strong red asymmetry, and peaks of both lines are red shifted for about 1000 km s$^{-1}$. 
We measured the full width half maximum (FWHM) of H$\beta$ of 5610 km s$^{-1}$, which is larger than the FWHM of
H$\gamma$ (5060 km s$^{-1}$), while the FWHM of rms profile of H$\gamma$ (4740 km s$^{-1}$) is only slightly higher than H$\beta$
one (4520 km s$^{-1}$).  In Figure \ref{mean1} we compare the mean and rms profiles of  H$\gamma$ and H$\beta$ lines. The residuals of the narrow lines subtraction have been artificially corrected in the rms profiles for better comparison.
As it can be seen in Figure \ref{mean1}, the rms profile of H$\beta$ shows smaller changes in the blue wing, and in the red part of the line, while the rms of H$\gamma$ shows smaller variability in the far red wing. 
We compare the mean H$\gamma$ and H$\beta$ profiles (upper panel) and their rms (bottom panel)
in Figure \ref{mean2}. It is interesting that the rms profiles seem to be the same, while, the mean H$\beta$  has more extensive red wing than the mean H$\gamma$.  It seems that there is an additional emission in the far wing of the mean H$\beta$, which is clearly seen when the two mean profiles are subtracted (Figure \ref{mean2}, upper panel). We fitted a simple Gaussian through the difference of the mean spectra, that is shifted to 7090 km s$^{-1}$ and with a FWHM of 5810 km s$^{-1}$.
We measured that the ratio of the red to the blue part of FWHM is
3.8 for the mean H$\beta$ and 3.1 for the mean H$\gamma$, that confirmed the observed difference in the asymmetry of the mean profiles.

\section{Discussion}

We have presented and analyzed the photometric and spectral data for QSO E1821+643 obtained from the long-term monitoring 
(2003--2014 for photometry; 1990--2014 for spectroscopy). 
Here we discuss the obtained results.

\subsection{Variation in the continuum and broad lines}

The results of the multi-wavelength monitoring of E1821+643 in X-ray, UV and optical ranges, simultaneously observed
for 37 days, were reported in \cite{ul92}. They found that there are no short-term changes in the UV and optical spectra,
while in the X-ray, there exists a variability on the short term scale. This result has been confirmed by \cite{ko93}, 
they also found possible changes on the larger time scale in the UV/optical spectra. Here, we explore the variability of
E1821+643 in the 24-year period in the continua at 4200\AA\, and 5100 \AA, and
in the broad H$\gamma$ and H$\beta$ emission lines. We find significant changes in the line and continuum 
fluxes during the monitored period. It is interesting that the H$\gamma$ line flux has changed for about two times (similar to continua at 4200 and 5100 \AA ), while the H$\beta$ line showed smaller variations in the line flux (1.4 times). 
We found that variations of both lines well correlate with variations of the corresponding continuum, but the 
response of the H$\gamma$ line is better than of the H$\beta$ line. 
This, and also the CCF analysis, indicates that the emitting region of the H$\beta$ line is distinctly larger than the emitting region of the H$\gamma$ line. This indicates a possible stratification in the BLR, showing smaller H$\gamma$ emitting region, that is also observed in a number of AGNs \citep[see Table 13 in][]{be10}. The BLR photoionization models predict a very similar equivalent distribution as a function of ionization parameter and density \citep{ko97}, i.e. one cannot expect a significant radial stratification in the broad Balmer lines emission regions. However, the possible radial stratification observed between H$\gamma$ and H$\beta$ emission regions in E1821+643 may be caused by optical-depth effects within the Balmer series \citep{kg04}. \cite{kg04} showed that there is a modest increase in responsivity between H$\beta$ and H$\gamma$ (see their Table 1). More detailed investigation of this is outside the scope of this paper.

The variability in the H$\beta$ line-segments is present, and it can be 
seen in Figure \ref{wings_c} that the line core and red wing fluxes are well correlated with the continuum flux at 5100 \AA, while a week correlation is present between the blue wing and continuum fluxes.

\subsubsection{Periodicity in the variability}

First we analyzed the photometric light curves and found  flare-like events of different amplitudes 
(0.1-0.5 magnitude), where the time interval between two consecutive maxima is $\sim$1000 days 
(see Figure \ref{fig1} and Table \ref{flare}).  
Similar maxima (1--3) are seen in the continuum light curves and possibly, with some delay of $\sim$100 days, in the H$\beta$ and H$\gamma$ line light curves, and in the broad H$\beta$ component and its line-segment light curves (see Figures \ref{lc} and \ref{lc-seg}). We note that it is difficult to detect these maxima in the observed spectroscopic light curves, given the scatter and sampling of data, however,
these shorter time-scale fluctuations become clearer in the GPR light curves shown in Figure \ref{gpr}.
Since it can indicate the periodical or quasi-periodical changes in the flux, we investigated the  
light curve periodicities and found
in all spectral light curves (in the continua and lines) the significant periodicity with the period of 4500 days. In the photometric light curves, there are three periods of 1200, 1850, and 4000 days. However,  as we noted above, the periodicity around 4000--4500 days obtained in spectroscopic and photometric light curves seems to cover only  one cycle in our observations and should be confirmed taking a longer observation campaign. These periods in the flux variability can indicate some kind of periodical rotation of some structures (in the disk around SMBH, or some cloudy-like gas) around the central SMBH.

\subsection{The SMBH mass and structure of the BLR}

The SMBH mass ($M_{BH}$) of E1821+643 can be estimated by using the virial theorem \citep[see][]{pet98,wa99}:

$$M_{\rm GRAV}=f{\Delta V_{\rm FWHM}\cdot R_{\rm BLR}\over G},$$
where $\Delta V_{\rm FWHM}$ is the  orbital velocity at that radius of the BLR, $R_{\rm BLR}$, and it is estimated from the width
 of the variable part of the  H$\beta$ emission line; $f$ is a factor that depends on the geometry of
the BLR and can be taken as $f=5.5$ \citep[][]{on04}. 
Taking into account that the dimension of the H$\beta$ BLR is 120 light days (see Table \ref{ccf}) and that the
FWHM of the  H$\beta$ rms profile is 4520 km s$^{-1}$, we obtained that the central SMBH has a mass of $2.6 \times10^9\ M_\odot$,
which is in agreement with estimates given by \cite{ko93}, they found that the mass is  $3 \times 10^9 M_{\odot}$.

To discuss the structure of the BLR, we should take into account that the broad line profiles have an unusual shape.
As we noted in section 2,  the broad line profiles  have a red asymmetry, i.e. the red wing is two 
times wider than the blue wing. In addition, the center of the broad H$\beta$ line component is red shifted for
$\sim$(1000$\pm$250) km s$^{-1}$ relatively to the peak of the narrow component. In the monitored period the peak 
position of the broad H$\beta$ component is varying  from $\sim$700km s$^{-1}$ to $\sim$1600 km s$^{-1}$ measured
as the centroid at 90\% of the maximal intensity (see Figure \ref{shift}). 
Note here that \cite{la08} reported similar redshifts, of 1000 -- 2000 km s$^{-1}$ for the broad components of H$\alpha$, H$\beta$ and several Paschen lines, and here we found that this redshift is changing.  

It is interesting that the broad H$\beta$ and H$\gamma$ components have different profiles (see Figure
\ref{mean2}), with the second one being narrower, since the H$\beta$ has an extended red wing. Additionally, the CCF analysis 
shows that the H$\gamma$  emitting region is significantly smaller ($\sim60$ days) than the H$\beta$ emitting region ($\sim 120$ days).
However, the shape of the rms of both lines are practically the same (see Fig. \ref{mean2}) that indicates a region which is variable, and that the emission from this region is mostly contributing to the line core of both lines.
The far H$\beta$ wing is emitted from another region that shows a smaller variability (see Fig. \ref{mean2}).

Possible shifts of the broad H$\beta$ line peak could be explained with a binary black hole model \citep[][]{pop12}
or with an inflowing  BLR \citep[][]{ga09}. It is not safe to conclude that an AGN is a supermassive black hole binary 
system (SMBB) based only on the broad line profiles, since the complex line profiles may be also caused by a complex BLR structure. 
However, as it was noted in \cite{pop12}, { the unusual broad line profiles together with 
other observational effects}, e.g. quasi-periodical oscillations or indications observed in  
spectropolarimetry, could be used for the SMBB detection. The merger hypothesis for E1821+643 has been discussed, 
and here we consider some results from spectral variability in the frame of this hypothesis.

\subsubsection{Recoiling supermassive black hole or supermassive black hole binary system}

A high redshifted broad emission lines, in the case of a SMBB system,
may be due to the emission from one BLR. { In this case, in the center of the circumbinary disk, the system makes a hole and the secondary SMBH orbits closer to the gas reservoir, and there is only one BLR}, i.e.  the best probability is that the smaller SMBH has only a BLR \citep[see e.g.][]{cu09}. In this case, we can expect higher radial velocities, as it is seen in the case of broad lines. However, problems with this scenario are that the distances between the two components have to be small \citep[see Table 1 in][]{pop12}, and the orbiting period should be shorter than the monitored period, and one can expect high changes in the shift (even one can observe a blueshift in the lines). Despite we detected some changes in the redshift of the broad H$\beta$ component (see Fig. \ref{shift}), they do not represent a dramatic change in the broad line shift. This scenario seems to be unlikely.
    
The second scenario is that the central SMBH and gas are a result of reminiscent of a previous interaction, the so called, recoiling SMBH. This is the case when a SMBH is fueled by the gas from the (former) 
gaseous (circumbinary) disk which falls into the SMBH \citep[see e.g.][]{za10}. 
 { In general, the interaction of the kicked SMBH with the interstellar medium is quite complicated, but simply the reprocessing of the X-ray emission could be responsible for the observed strong emission lines. 
    In that case, broad lines originate from a standard BLR associated to the recoiling SMBH,} 
    and narrow lines are associated to the host 
    galaxy. { The offset in broad emission lines could be detected directly after a high-velocity recoil or at the time of pericentric passages through a gas-rich remnant.} The shift of a broad line (with respect 
    to the narrow one) can be expected, and kick velocities can be of an order of 1000 km s$^{-1}$, i.e.
    there is a probability that 23\% of recoils are larger than 1000 km s$^{-1}$ \citep[see][]{lo10}. 
    This scenario is in agreement with the observed velocities in the broad lines of E1821+643. However, one 
    should take into account the line-of-sight projection of the velocity, that will always give smaller projected velocities.
 
\cite{ro10} analyzed the spectropolarimetric observations and showed that the E1821+643 spectrum 
is only weakly polarized, with an average degree of polarization of 0.21\% $\pm$ 0.03\% at a position angle 140$\circ \pm 5\circ$. The average polarization position angle is approximately perpendicular to the arcsecond-scale radio source. They found  that  in the polarized flux the broad H$\alpha$ line shows a strong blue asymmetry and   a 
similar ($\sim$1000 km s$^{-1}$) blueshift of the peak.
{ \cite{ro10} considered if a possible explanation of their observations could be the scattering of the broad emission lines coming from the active component of an SMBH binary, or the outflowing wind.} For an SMBB system there is a problem with the polarized angle,
i.e. in this case the { scattering geometry would produce the polarization aligned with the direction of the radio jet, which is in contrast to the observations.} 
They  support an interpretation of this results  in the framework 
of the hypothesis of a recoiling SMBH.  They found that the SMBH is itself 
moving with a velocity $\sim$2100 km s$^{-1}$ relative to the host galaxy. However, to accept the recoil hypothesis, 
it seems there should be a very specific  coalescence binary configuration.

Additionally, we have to note here that we found some periodicities in the variability of photometric and 
spectral data, that also may be connected with a binary system. Considering the recoiling scenario,
one has a problem to explain the periodicity detected in the flux variability and also a huge kick-off velocity 
of the SMBH. However, the extra-nuclear gas indicates tidal interaction or merger process in the center of
E1821+643 \citep{fr98,ar11}, but it is not clear if the source of detected gas came from  a
gas-rich companion galaxy that is merging with the quasar elliptical host galaxy or it is a 
reminiscent of a previous  collision. In fact, the periodicity in the flux variation, may
be caused by orbiting of very dense gas-rich cloudy-like structures \citep[see][]{ar11}
around recoiling SMBH. We hope that a detailed investigation of the 
line profile variability (planned in Paper II) will give more information about the structure of the BLR.
Once again we should point out that the observed asymmetry of broad lines, as well as in
spectrophotometric observations could be explained by the complex BLR geometry \citep[see][]{ga09,pop12}.

\section{Conclusion}

We have presented a long-term photometric and spectrophotometric monitoring campaign for E1821+643.
The photometric data  for  2003--2014 period (98 nights) are presented in a photometric system close to the Johnson (BV-filter) and Cosin (R filter) systems. The spectral data for 1990--2014 period (127 spectra in H$\beta$ and 76 spectra in H$\gamma$) were unified by the absolute scaling of the observed spectra to the flux of [OIII]4959+5007 lines and are corrected for aperture effects.

\begin{figure} 
\centering
\includegraphics[width=13cm]{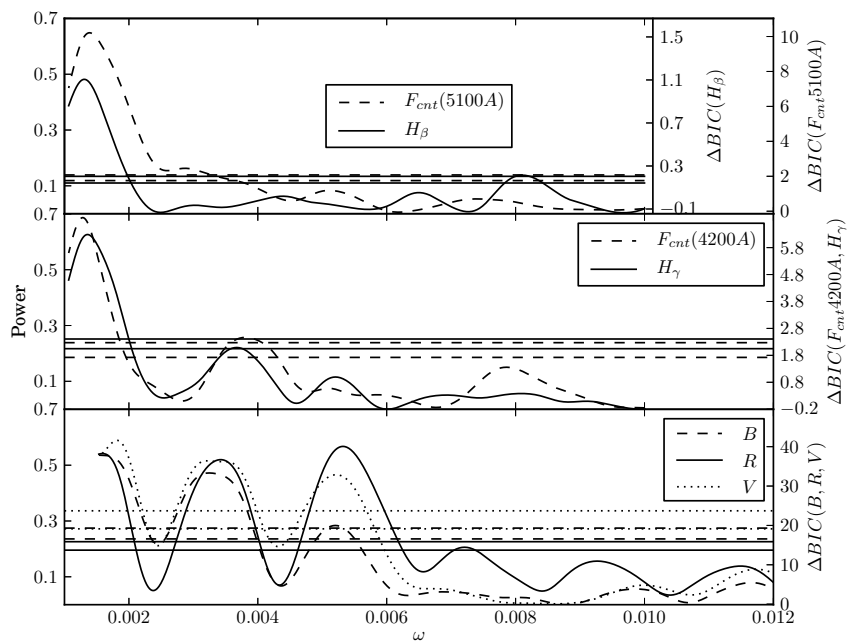}
\caption{A comparison of GLS for spectrophotometric light curves.
Horizontal lines show the 1\% and 5 \% significance levels for the
 highest peak determined by 1000 bootstrap samplings.  The difference  in
Bayesian information criterion ($\Delta BIC$) compares
 single harmonic model and pure noise model. X axis depicts angular
frequencies $\omega=\frac{2\pi}{period}$.} \label{period1} 
\end{figure}

\begin{figure}
\centering
\includegraphics[width=11cm]{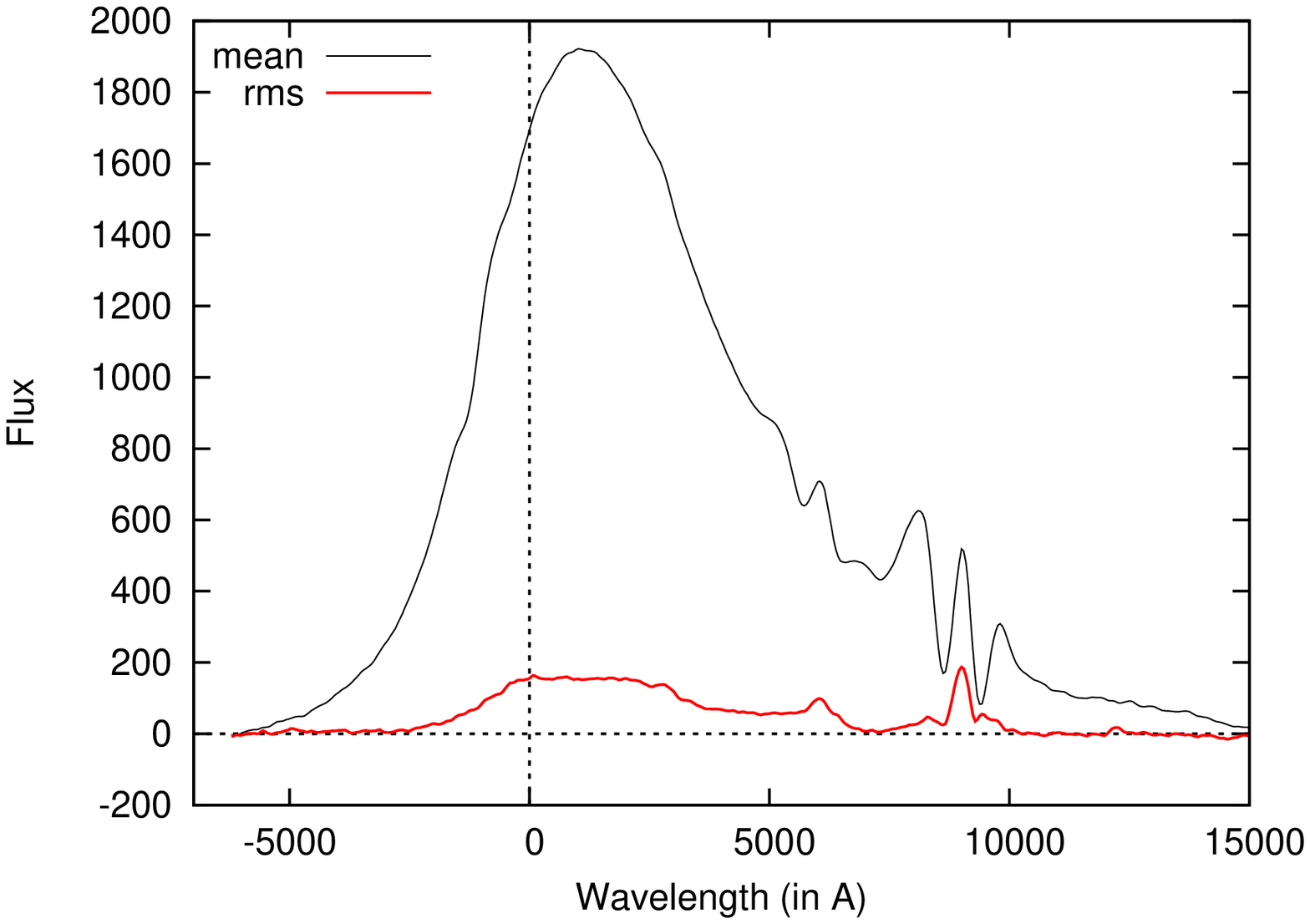}
\includegraphics[width=11cm]{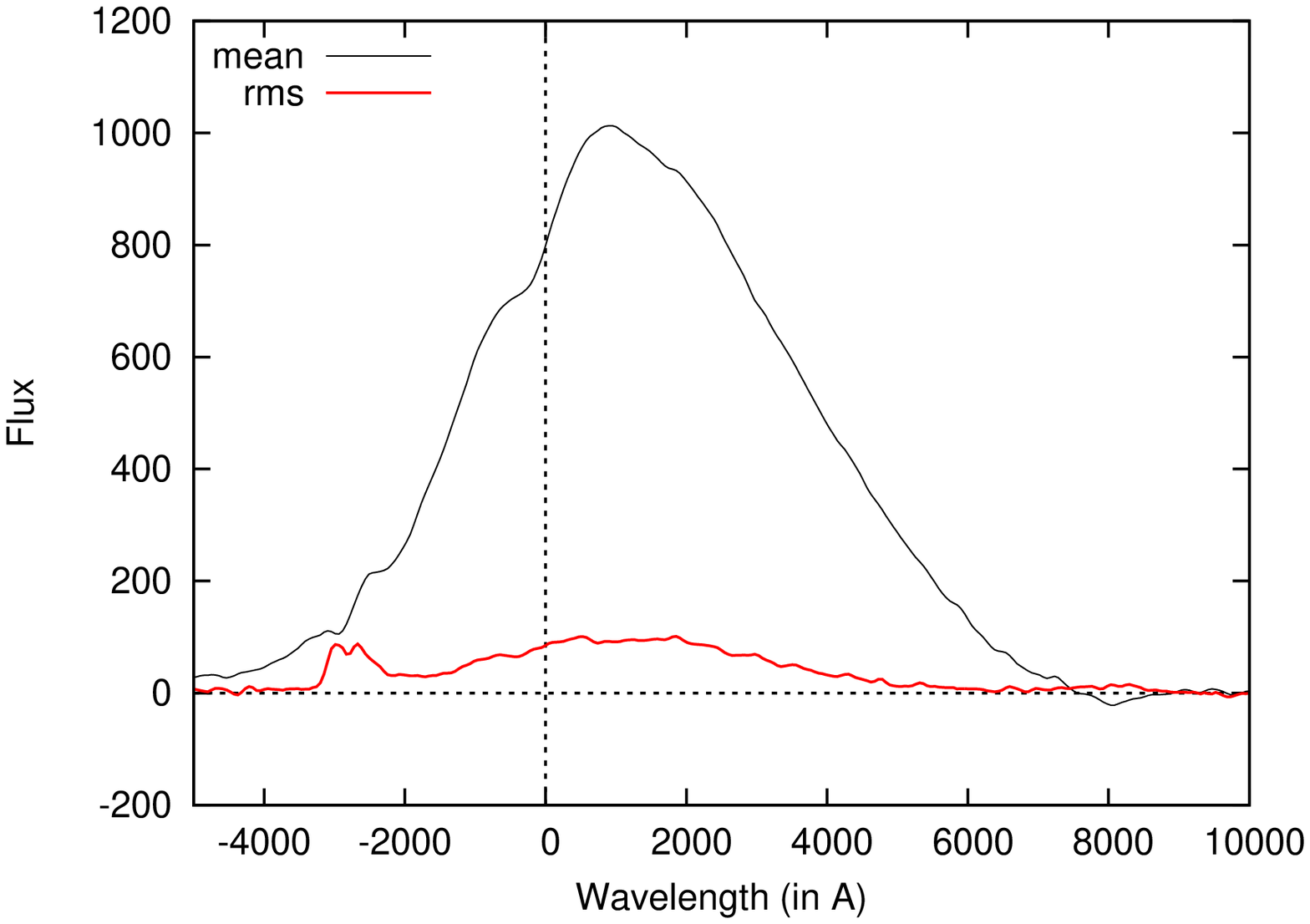}
\caption{Mean and rms spectra of the broad H$\beta$ (upper) and H$\gamma$ lines (bottom). } \label{mean}
\end{figure}

\begin{figure}
\centering
\includegraphics[width=11cm]{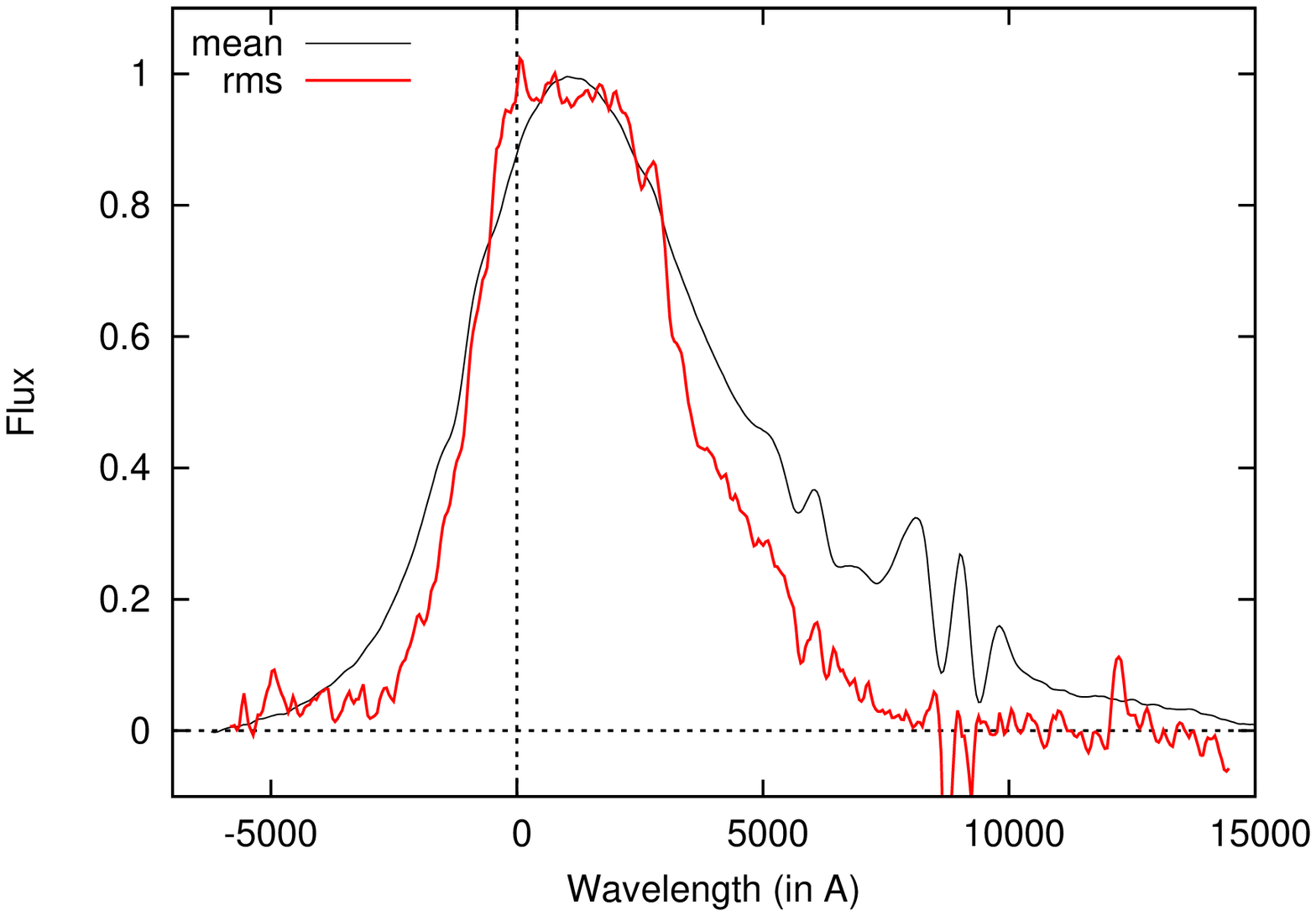}
\includegraphics[width=11cm]{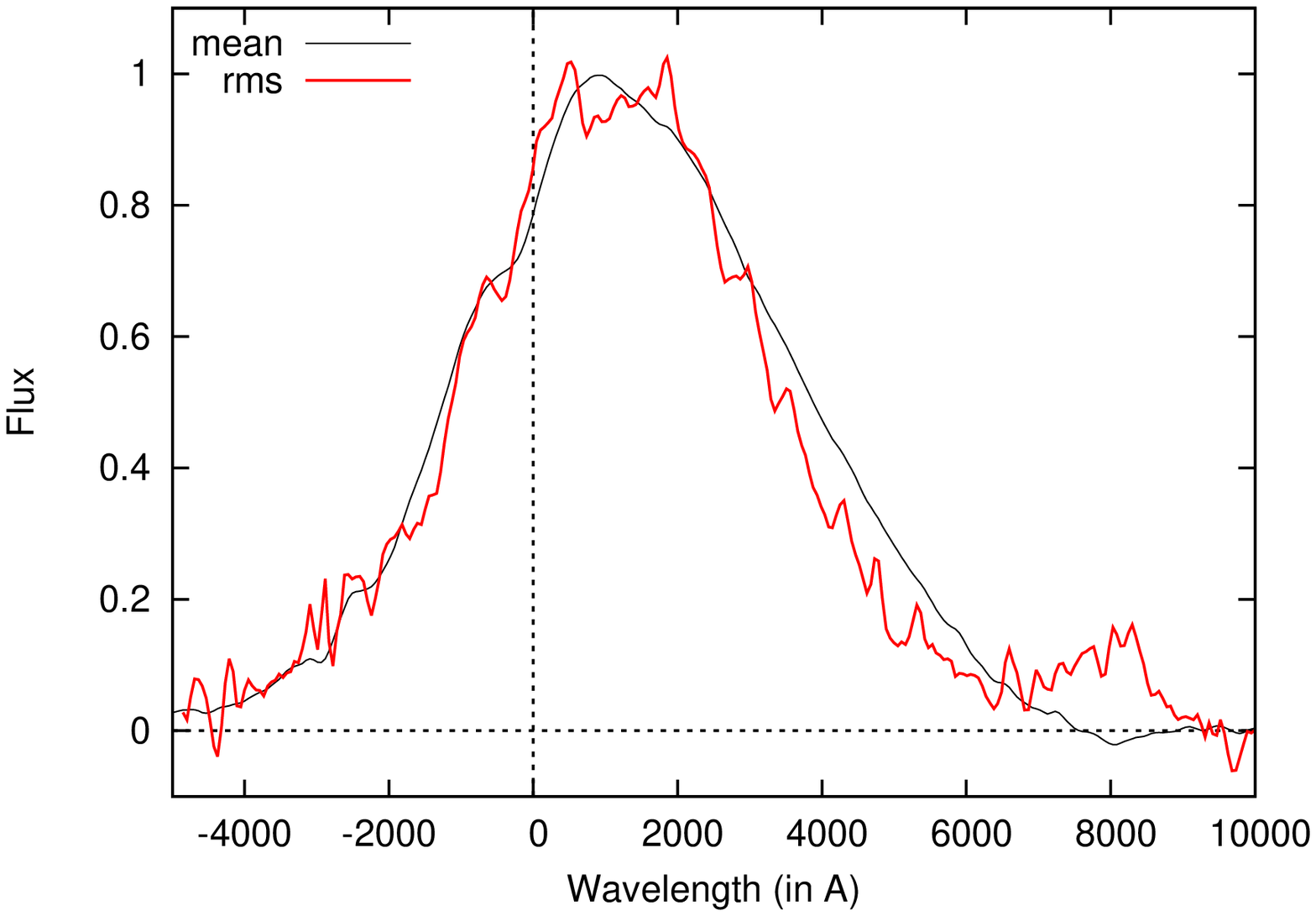}
\caption{The comparison of mean and rms spectra of the broad H$\beta$ (upper) and H$\gamma$ (bottom). } \label{mean1}
\end{figure}

\begin{figure}
\centering
\includegraphics[width=11cm]{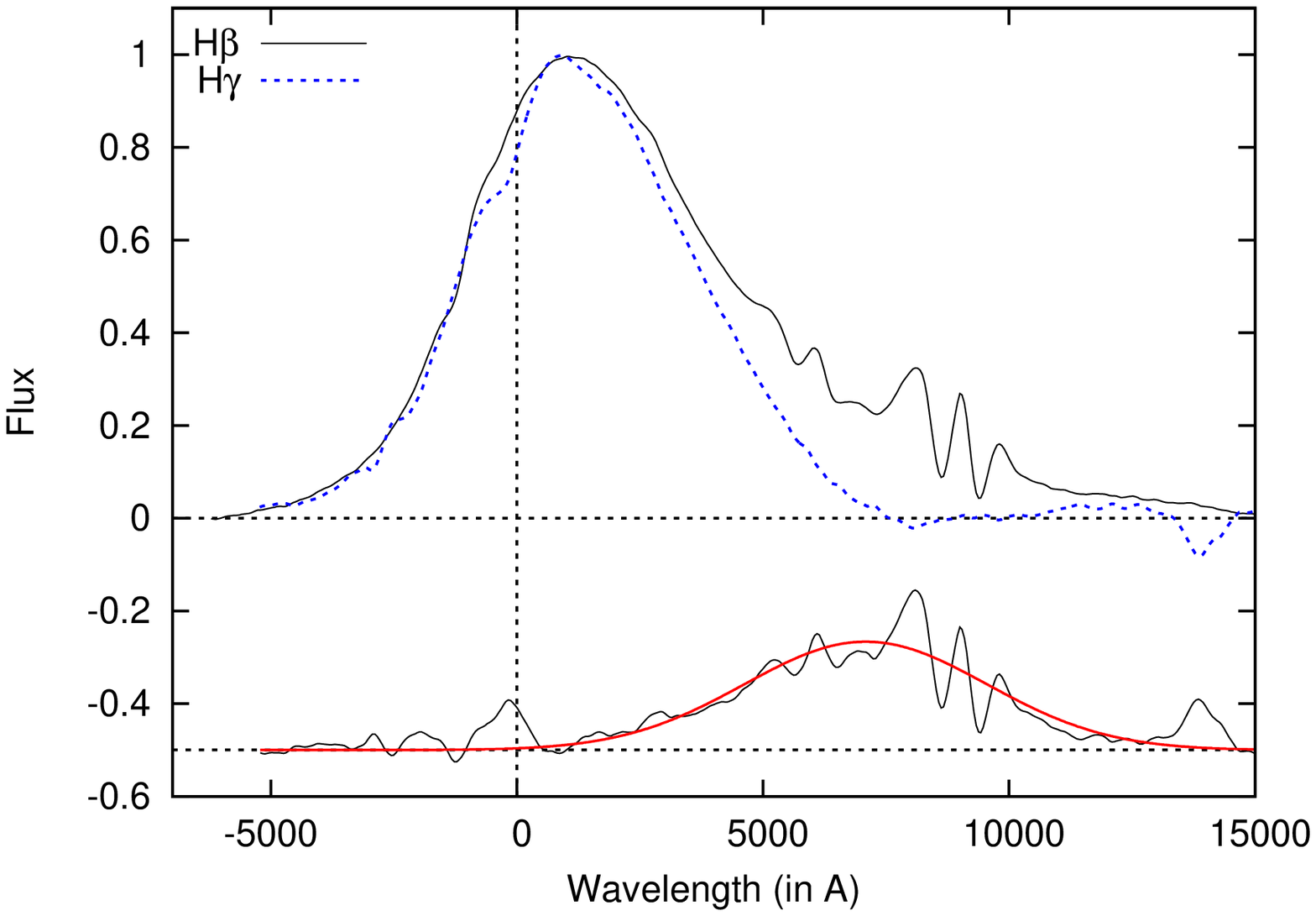}
\includegraphics[width=11cm]{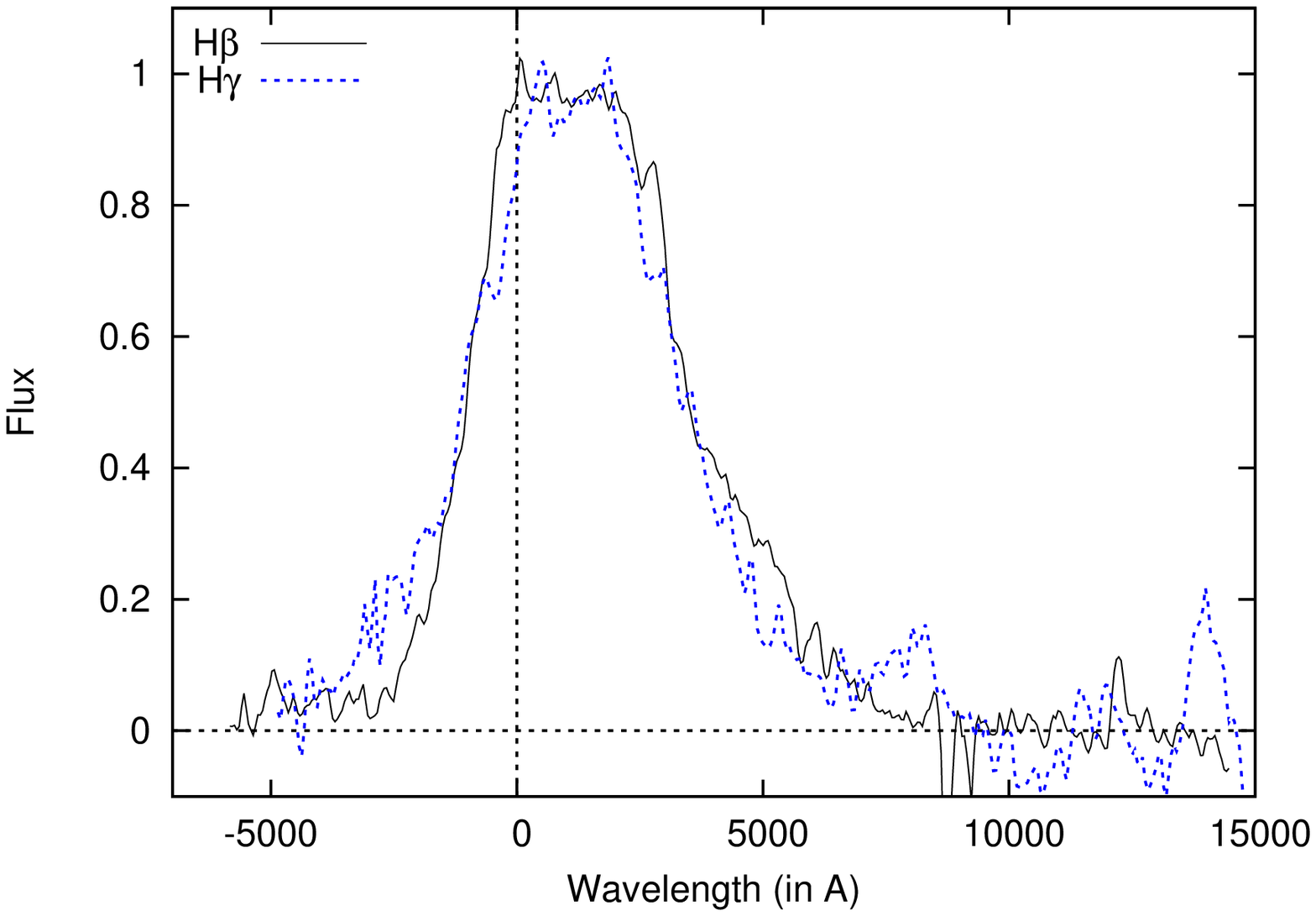}
\caption{The comparison of the mean broad H$\beta$ and H$\gamma$ profiles (upper) and their rms (bottom). The difference between the 
H$\beta$ and H$\gamma$ mean profiles is also given at the bottom of the upper plot, fitted with a single Gaussian.} \label{mean2}
\end{figure}

\begin{figure*}
\centering
\includegraphics[width=7cm,angle=-90]{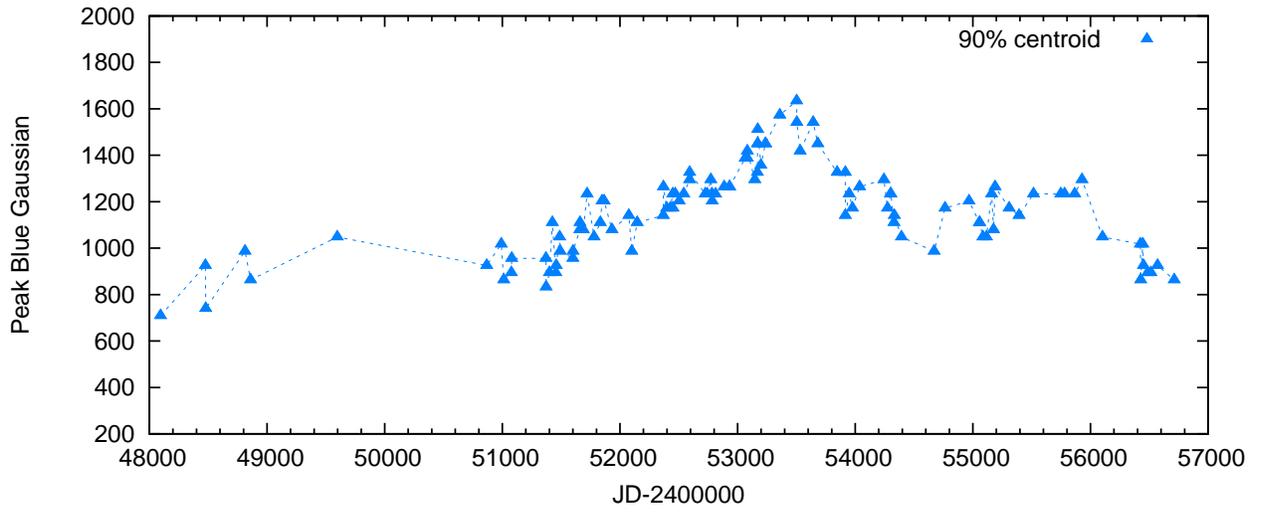}
\caption{The changes in the redshift of the broad H$\beta$ component during the monitored period, measured
as the centroid at 90\% of the maximal intensity.} \label{shift}
\end{figure*}

We have constructed the continua, H$\beta$ and H$\gamma$ lines light curves and  investigated the flux variations in the continua and in the total H$\beta$ and H$\gamma$ line fluxes, as well as in the broad H$\beta$ line-segment fluxes. We have cross correlated the continuum and broad line fluxes and investigated the periodicity in the photometric and spectral flux variation. From our investigation we can outline the following conclusions:

\begin{enumerate}

\item The fluxes in the continuum (at 5100 \AA\ and 4200 \AA\ in the rest frame) and total H$\gamma$ line varied for about 2 times, while the total H$\beta$ line flux varied for around 1.4 times (see Table \ref{var}) during the monitored period. The amplitude of variability F(var) is $\sim$19\% for continua and total H$\gamma$ line, while it is  $\sim$7\%  for the total  H$\beta$ line. This may be caused by different dimension of the
H$\gamma$ and H$\beta$ emitting region, since our CCF analysis shows that the H$\gamma$ emission region is significantly (two times) smaller than the H$\beta$ one. The CCF of the continuum at 5100 \AA\ and total H$\beta$ 
emission line fluxes shows a lag of $\sim$120 days, while the lag
between  the continuum at 4200 \AA\  and total H$\gamma$ line  is $\sim$60 days. The CCF between the continua at
4200 \AA\ and  at 5100 \AA\ shows a short time lag of $\sim$2-6 days. This difference in
the broad line response delays to the corresponding continuum changes indicates some kind of stratification in the BLR. 
However, the H$\beta$ and H$\gamma$ line fluxes and the H$\beta$ broad line-segments are well correlated 
with the continuum flux, indicating that the BLR of E1821+643 is primarily photoionized by the
central continuum source and the  ionizing continuum is a good extrapolation of the optical continuum.

\item The broad Balmer lines have extremely asymmetric profiles, with the red wing extending to Doppler velocities of at least $\sim$15000 km s$^{-1}$ relative to the rest frame wavelength. The peak of the broad H$\beta$ component was red shifted relative to the corresponding narrow lines, between  $\sim$700  and 1600 km s$^{-1}$  during the monitored period (see Figure \ref{shift}).  We found that the mean broad  H$\gamma$ and H$\beta$ profiles are different, with the H$\beta$ showing an extensive red wing that is broader than H$\gamma$ red wing. However, the rms of both lines are practically the same, showing that the line core is the most variable component. We used the estimated BLR dimension and FWHM of the H$\beta$ rms profile to find the mass of SMBH. We estimate that the SMBH mass in E1821+643 is $2.6 \times10^9\ M_\odot$, that is in a good agreement with earlier estimations from the Balmer bump.

\item  During 2003--2014 the photometric continuum fluxes in the BVR filters  has showed  an almost sinusoidal change ($\sim$4 maxima or flare-like events), that motivated us to explore whether any periodicity is present in the photometric and spectroscopic light curves. In the photometric light curves we found  three periods of variability: 4000, 1850, and 1200 days, while in the spectroscopic variability (in broad lines and continuum) we found one significant period of 4500 days (see Figs. \ref{period1}). Note here that the periodicity of 4000--4500 days is observed as a cycle periodicity and should be taken with caution. The periodicity of the light curves is probably connected with  an orbital motion around the central black hole. In this case, it is hard to explain this fact with the  binary black hole hypothesis where only one BLR is present, because the shift of the broad lines stays always in the red part of the line. However, there may be a possibility that the periodical variability is caused by gas-rich cloudy-like structures which are orbiting around the recoiling black hole.

\end{enumerate}

We are going to investigate the broad line profiles in more details in Paper II and to discuss the changes in the spectral energy distribution, and the contribution of the Balmer continuum and changes in Balmer bump.

\acknowledgments

This work was supported by: INTAS (grant N96-0328), RFBR (grants
N97-02-17625 N00-02-16272, N03-02-17123, 06-02-16843, N09-02-01136,
12-02-00857a, 12-02-01237a,N15-02-02101), CONACYT research grants 39560, 54480,
and 151494, and the Ministry of Education and Science of Republic of Serbia through the project
Astrophysical Spectroscopy of Extragalactic Objects (176001). 
L. \v C. P., W. K. and D. I. are grateful to the Alexander von Humboldt
foundation for support in the frame of program "Research Group
Linkage". W. K. is supported by the DFG Project Ko 857/32-1. 
D.I. has been awarded L'Or\'eal-UNESCO “For Women in Science” National Fellowship for 2014. 
We especially thank Borisov N.V., Moiseev A., and Vlasuyk V.V. for taking part in 
the observations.

%\bibitem [Alexander(1997)]{al97} Alexander, T. 1997, Astronomical Time Series, ed. D. Maoz, A. Sternberg, \& E. M. Leibowitz 
%(Dordrecht: Kluwer), 163
%\bibitem[Bischoff \& Kollatschny(1999)]{bi99} Bischoff, K., \& Kollatschny, W. 1999, \aap, 345, 49B
%\bibitem[Boller et al.(1996)]{bol96}   Boller, Th., Brandt, W. N.,  Fink, H. 1996, \aap, 305, 53
%\bibitem[Boroson \& Green(1992)]{bo92}  Boroson, T. A., \& Green, R. F. 1992, \apjs, 80, 109

\clearpage

%%%%%%%%%%%%%%%%%%%%%%%%%%%%%%%%
%%%%%%%%%%%%%%%%%%%%%%%%%%%%%%%%

\clearpage

%%%%%%%%%%%%%%%%%%%%%%%%%%%%%%%%
%\begin{figure*}
%\centering
%\includegraphics[width=8cm]{ph_sp.ps}
%\includegraphics[width=8cm]{ph_sp-V.ps}
%\caption{Correlations between the photometric and spectral fluxes in the monitored 
%period: F(R) vs. F(5100) (left) and F(V) vs. F(4200) (right).  
%The correlation coefficients and the corresponding p-values are indicated in the bottom right corner.
%The error bars of photometric fluxes are comparable to the symbol size.} \label{fig2}
%\end{figure*}
%%%%%%%%%%%%%%%%%%%%%%%%%%%%%%%%

\clearpage

%%%%%%%%%%%%%%%%%%%%%%%%%%%%%%%%
%%%%%%%%%%%%%%%%%%%%%%%%%%%%%%%%

\clearpage

%%%%%%%%%%%%%%%%%%%%%%%%%%%%%%%%
%%%%%%%%%%%%%%%%%%%%%%%%%%%%%%%%

\clearpage

%%%%%%%%%%%%%%%%%%%%%%%%%%%%%%%%
%%%%%%%%%%%%%%%%%%%%%%%%%%%%%%%%

\clearpage

%%%%%%%%%%%%%%%%%%%%%%%%%%%%%%%%
%%%%%%%%%%%%%%%%%%%%%%%%%%%%%%%%

\clearpage

%%%%%%%%%%%%%%%%%%%%%%%%%%%%%%%%
%%%%%%%%%%%%%%%%%%%%%%%%%%%%%%%%

\clearpage

%%%%%%%%%%%%%%%%%%%%%%%%%%%%%%%%
%%%%%%%%%%%%%%%%%%%%%%%%%%%%%%%%

\clearpage

%%%%%%%%%%%%%%%%%%%%%%%%%%%%%%%%
%%%%%%%%%%%%%%%%%%%%%%%%%%%%%%%%

\clearpage

%%%%%%%%%%%%%%%%%%%%%%%%%%%%%%%%
%%%%%%%%%%%%%%%%%%%%%%%%%%%%%%%%

\clearpage

%%%%%%%%%%%%%%%%%%%%%%%%%%%%%%%%
%%%%%%%%%%%%%%%%%%%%%%%%%%%%%%%%

\clearpage

%%%%%%%%%%%%%%%%%%%%%%%%%%%%%%%%
%%%%%%%%%%%%%%%%%%%%%%%%%%%%%%%%

\clearpage

%%%%%%%%%%%%%%%%%%%%%%%%%%%%%%%%
%%%%%%%%%%%%%%%%%%%%%%%%%%%%%%%%

\clearpage

%%%%%%%%%%%%%%%%%%%%%%%%%%%%%%%%
%%%%%%%%%%%%%%%%%%%%%%%%%%%%%%%%

\clearpage

%%%%%%%%%%%%%%%%%%%%%%%%%%%%%%%%
%%%%%%%%%%%%%%%%%%%%%%%%%%%%%%%%

\clearpage

%%%%%%%%%%%%%%%%%%%%%%%%%%%%%%%%
%\begin{figure} 
%\centering
%\includegraphics[width=10cm]{period.eps}
%\caption{Best fit to the spectrophotometric curves using sum of cosines
%and sinus functions, with angular frequencies selected using GLS. Note that in the case of H$\beta$ and H$\gamma$ we added angular frequencies corresponding to 1300 days period.} \label{period2} 
%\end{figure}
%%%%%%%%%%%%%%%%%%%%%%%%%%%%%%%%

\clearpage

%%%%%%%%%%%%%%%%%%%%%%%%%%%%%%%%
%%%%%%%%%%%%%%%%%%%%%%%%%%%%%%%%
\clearpage

%%%%%%%%%%%%%%%%%%%%%%%%%%%%%%%%
%%%%%%%%%%%%%%%%%%%%%%%%%%%%%%%%

\clearpage

%%%%%%%%%%%%%%%%%%%%%%%%%%%%%%%%
%%%%%%%%%%%%%%%%%%%%%%%%%%%%%%%%

\clearpage

%%%%%%%%%%%%%%%%%%%%%%%%%%%%%%%%
%%%%%%%%%%%%%%%%%%%%%%%%%%%%%%%%

\clearpage

%Table Photometric Flux
%%%%%%%%%%%%%%%%%%%%%%%%%%%%%%%%
\begin{deluxetable}{clccccc}
\tabletypesize{\scriptsize}
\tablecaption{The measured photometric magnitudes.\label{tab1}}
\tablewidth{0pt}
\tablehead{
\colhead{N} & \colhead{UT-date} & \colhead{MJD} & \colhead{Seeing} & \colhead{M$_{\rm R}$ $\pm \sigma$} &
\colhead{M$_{\rm V}$ $\pm \sigma$} & \colhead{M$_{\rm B}$ $\pm \sigma$}  \\
  &    & \colhead{2400000+} & \colhead{[arcsec]} &  &    &   \\
\colhead{1} & \colhead{2} & \colhead{3} & \colhead{4} & \colhead{5} & \colhead{6} & \colhead{7}
}
\startdata
  1  & 28.06.03 & 52818.4 &  3.0  & 13.964 $\pm$ 0.027   &  15.230 $\pm$ 0.030  & 14.413 $\pm$ 0.015\\
  2  & 30.06.03 & 52820.5 &  3.0  & 13.965 $\pm$ 0.036   &  15.212 $\pm$ 0.028  & 14.431 $\pm$ 0.014\\
  3  & 31.07.03 & 52852.4 &  2.0  & 13.956 $\pm$ 0.028   &  15.232 $\pm$ 0.008  & 14.420 $\pm$ 0.009\\
  4  & 26.08.03 & 52877.4 &  2.3  & 13.980 $\pm$ 0.036   &  15.276 $\pm$ 0.029  & 14.430 $\pm$ 0.026\\
  5  & 01.10.03 & 52913.4 &  1.8  & 13.967 $\pm$ 0.036   &  15.283 $\pm$ 0.014  & 14.443 $\pm$ 0.029\\
  6  & 02.10.03 & 52914.4 &  1.8  &  -                   &  15.265 $\pm$ 0.035  & 14.462 $\pm$ 0.029\\
  7  & 04.12.03 & 52978.2 &  2.0  & 13.978 $\pm$ 0.032   &  15.285 $\pm$ 0.035  & 14.470 $\pm$ 0.025\\
  8  & 22.04.04 & 53117.5 &  2.2  & 13.960 $\pm$ 0.027   &  15.205 $\pm$ 0.014  & 14.405 $\pm$ 0.022\\
\enddata
\tablecomments{Col.(1): Number. Col.(2): UT date. Col.(3): Modified Julian
date (MJD). Col.(4): Mean seeing in arcsec. Cols.(5)-(7): BRV magnitudes
and the corresponding errors.}
\end{deluxetable}
%%%%%%%%%%%%%%%%%%%%%%%%%%%%%%%%
\clearpage

%%%%%%%%%%%%%%%%%%%%%%%%%%%%%%%%
\begin{deluxetable}{lcccc}
\tabletypesize{\scriptsize}
\tablecaption{Sources of spectroscopic observations.\label{tab2}}
\tablewidth{0pt}
\tablehead{
\colhead{Observatory} & \colhead{Code} & \colhead{Tel.aperture + equipment} & \colhead{Aperture [arcsec]} & \colhead{Focus} \\
\colhead{1} & \colhead{2} & \colhead{3} & \colhead{4} & \colhead{5}
}
\startdata
SAO (Russia)                  & L(N)    & 6 m + Long slit  &  2.0$\times$6.0   & Nasmith    \\
SAO (Russia)                  & L(U)    & 6 m + UAGS       &  2.0$\times$6.0   & Prime      \\
SAO (Russia)                  & L(Sc)   & 6 m + Scorpio    &  1.0$\times$6.07  & Prime      \\
Guillermo Haro Observatory (M\'exico)     & GHO     & 2.1 m + B\&C     &  2.5$\times$6.0   & Cassegrain   \\
SAO (Russia)                  & Z1      & 1 m + GAD        &  4.0$\times$19.8  & Cassegrain   \\
SAO (Russia)                  & Z2      & 1 m + GAD        &  4.0$\times$9.45  & Cassegrain   \\
Calar Alto(Spain)             & CA1     & 3.5 m + B\&C / TWIN &  2.0$\times$4.0 &   Cassegrain \\
Calar Alto(Spain)             & CA2     & 2.2 m + B\&C        &  2.0$\times$4.0 &   Cassegrain \\
\enddata
\tablecomments{Col.(1): Observatory. Col.(2): Code assigned to each combination of 
telescope + equipment used throughout this paper. Col.(3): Telescope aperture and spectrograph. 
Col.(4): Projected spectrograph entrance apertures (slit width$\times$slit length in
arcsec). Col.(5): Focus of the telescope.\\}
\end{deluxetable}
%%%%%%%%%%%%%%%%%%%%%%%%%%%%%%%%

%%%%%%%%%%%%%%%%%%%%%%%%%%%%%%%%
\begin{deluxetable}{clccccc}
\tabletypesize{\scriptsize}
\tablecaption{The log of spectroscopic observations.\label{tab3}}
\tablewidth{0pt}
\tablehead{
\colhead{N} & \colhead{UT-date} & \colhead{MJD} & \colhead{CODE\tablenotemark{a}} & \colhead{Aperture} &
\colhead{Sp.range} & \colhead{Seeing}  \\
  &    & \colhead{2400000+} &  & \colhead{[arcsec]} & [${\rm \AA}$]   &  \colhead{[arcsec]} \\
\colhead{1} & \colhead{2} & \colhead{3} & \colhead{4} & \colhead{5} & \colhead{6} & \colhead{7}
}
\startdata
    1  & 23.07.1990  &48095.0 & CA1   & 2.0x4.0 &6060-7690  & 1.5 - 2.5  \\
    2  & 07.08.1991  &48476.0 & CA2   & 2.0x4.0 &5170-7650  & 1.5 - 2.5 	\\
    3  & 10.08.1991  &48479.0 & CA2   & 2.0x4.0 &5540-7640  & 1.5 - 2.5 	\\
    4  & 09.07.1992  &48813.0 & CA1   & 2.0x4.0 &5510-7690  & 1.5 - 2.5 	\\
    5  & 25.08.1992  &48860.0 & CA2   & 2.0x4.0 &3850-7650  & 1.5 - 2.5 	\\
    6  & 01.09.1994  &49597.0 & CA2   & 2.0x4.0 &5970-7690  & 1.5 - 2.5 	\\
    7  & 22.02.1998  &50866.6 & L(N)  &2.0x6.0 &4690-6930  & 2     \\
    8  & 07.05.1998  &50941.1 & L(N)  &2.0x6.0 &4350-6630  & 3     \\
\enddata
\tablecomments{Col.(1): Number. Col.(2): UT date. Col.(3): Modified Julian
date (JD). Col.(4): CODE\tablenotemark{a}. Col.(5): Projected
spectrograph entrance apertures. Col.(6): Wavelength range covered.
Col.(7): Mean seeing in arcsec.}
\tablenotetext{a}{Code given according to Table~\ref{tab1}.}
\end{deluxetable}
%%%%%%%%%%%%%%%%%%%%%%%%%%%%%%%%  
  
\clearpage

\begin{deluxetable}{lccc}
%\tabletypesize{\scriptsize}
\tablecaption{Flux scale factors for optical spectra.\label{tab4}}
\tablewidth{0pt}
\tablehead{
\colhead{Sample} & \colhead{Aperture} & \colhead{Scale factor} & \colhead{Extended source  correction} \\
  &  \colhead{(arcsec)} & \colhead{($\varphi$)} & \colhead{G(g)\tablenotemark{a}}
}
\startdata
L(U,N)     &  2.0$\times$6.0   & 1.000  &  0.000 \\
L(Sc)      &  1.0$\times$6.07   & 0.976  & -0.414 \\
GHO        &  2.5$\times$6.0   & 0.951  &  0.000 \\
Z1         &  4.0$\times$19.8  & 0.958$\pm$0.017  &  0.280$\pm$0.166 \\
Z2         &  4.0$\times$9.45  & 1.048  &  0.033 \\
CA1        &  2.0$\times$4.0   & 1.000    & 0.000 \\
CA2        &  2.0$\times$4.0   & 1.000    & 0.000 \\
\enddata
\tablenotetext{a}{In units of 10$^{-15} \rm erg \ cm^{-2} s^{-1}$\AA$^{-1}$}
\end{deluxetable}
%%%%%%%%%%%%%%%%%%%%%%%%%%%%%%%%%%%%%%%%%%%%%%%%%%%%%%%%%%%%%%%%

\clearpage

%Table Fluxes
%%%%%%%%%%%%%%%%%%%%%%%%%%%%%%%%
\begin{deluxetable}{clccccc}
\tabletypesize{\scriptsize}
\tablecaption{The measured line and continuum fluxes.\label{tab5}}
\tablewidth{0pt}
\tablehead{
\colhead{N} & \colhead{UT-date} & \colhead{JD+} & \colhead{F${\rm 5100}\pm \sigma$} & \colhead{F(H$\beta$)$\pm \sigma$} &
\colhead{F${\rm 4200}\pm \sigma$} & \colhead{F(H$\gamma$)$\pm \sigma$}  \\
  &    & \colhead{2400000+} &  \colhead{$10^{-15} \rm erg \, cm^{-2} s^{-1}$\AA$^{-1}$} & 
\colhead{$ 10^{-13} \rm erg \, cm^{-2} s^{-1}$} & \colhead{$ 10^{-15} \rm erg \, cm^{-2} s^{-1}$}   
& \colhead{$ 10^{-13} \rm erg \, cm^{-2} s^{-1}$}   \\
\colhead{1} & \colhead{2} & \colhead{3} & \colhead{4} & \colhead{5} & \colhead{6} & \colhead{7} 
}
\startdata
    1  & 23.07.1990  &48095.0 &   5.23 $\pm$ 0.12 &    11.58 $\pm$  0.34   &   7.58 $\pm$    0.24  &     2.60 $\pm$ 0.14  \\  
    2  & 07.08.1991  &48476.0 &   5.59 $\pm$ 0.13 &    11.29 $\pm$  0.33   &   7.34 $\pm$    0.24  &     2.34 $\pm$ 0.12  \\  
    3  & 10.08.1991  &48479.0 &   5.42 $\pm$ 0.13 &    10.81 $\pm$  0.32   &   7.20 $\pm$    0.23  &     2.28 $\pm$ 0.12  \\ 
    4  & 09.07.1992  &48813.0 &   5.19 $\pm$ 0.12 &    11.65 $\pm$  0.34   &   6.50 $\pm$    0.21  &     2.39 $\pm$ 0.12  \\ 
    5  & 25.08.1992  &48860.0 &   5.50 $\pm$ 0.13 &    12.32 $\pm$  0.36   &  12.36 $\pm$    0.40  &     2.84 $\pm$ 0.15  \\   
    6  & 01.09.1994  &49597.0 &   6.31 $\pm$ 0.15 &    11.59 $\pm$  0.34   &        -              &         -           \\   
    7  & 22.02.1998  &50866.6 &   8.43 $\pm$ 0.20 &    12.55 $\pm$  0.37   &        -              &         -           \\   
    8  & 07.05.1998  &50941.1 &   8.76 $\pm$ 0.21 &    12.23 $\pm$  0.36   &  10.65 $\pm$   0.34   &     2.61 $\pm$ 0.14 \\
\enddata
\end{deluxetable}
%%%%%%%%%%%%%%%%%%%%%%%%%%%%%%%%

\clearpage

%%%%%%%%%%%%%%%%%%%%%%%%%%%%%%%%
\begin{deluxetable}{lcccccc}
%\tabletypesize{\scriptsize}
\tablecaption{Estimates of the mean errors for H$\beta$ and H$\gamma$ total-line fluxes  and 
for H$\beta$ broad-line segment fluxes.\label{tab6}}
\tablewidth{0pt}
\tablehead{
 \colhead{Line} &  \colhead{Spectral Region} &  \colhead{Spectral Region}     & \colhead{$\sigma \pm$e} & \colhead{$V_r$ region} & \colhead{$|V_{\rm red}-V_{\rm blue}|$}\\
    &  \colhead{[\AA] (obs)} & \colhead{[\AA] (rest)} & \colhead{[\%]} & \colhead{km s$^{-1}$} & \colhead{km s$^{-1}$}
}
\startdata
 cont 5100         &  6601-6631  & 5089-5111  & 2.3 $\pm$2.0  &  -  & - \\
 cont 4200         &  5459-5489  & 4208-4231  & 3.2 $\pm$2.1  &  -  & - \\
 H$\beta$ - total  &  6170-6620  & 4756-5103  & 2.9 $\pm$2.1  &  (-6423) - (+14988)  &  21411  \\
 H$\gamma$ - total &  5550-5770  & 4278-4448  & 5.2 $\pm$3.1  &  (-4763) - (+7460)   &  12223  \\
 H$\beta$ - broad  &  6170-6620  & 4756-5103  & 2.9 $\pm$2.1  &  (-6423) - (+14988)  &  21411   \\
 H$\beta$ - blue   &    6199-6284 & 4779-4844 & 4.9$\pm$3.8   & (-6089) - (-2091)   &   3998  \\
 H$\beta$ - core   &    6285-6369 & 4845-4910 & 3.1$\pm$3.0   & (-2030) - (+1968)   &   3998  \\
 H$\beta$ - red    &    6371-6456 & 4911-4977 & 4.1$\pm$3.2   & (+2030) - (+6088)   &   4058  \\
 H$\beta$ - far red &   6457-6551 & 4978-5050 & 6.0$\pm$4.6   & (+6150) - (+10578)  &   4428  \\
\enddata
\tablecomments{Col.(1): Measured continuum/line/line-segment. Col.(2): Observed wavelength range.
Col.(3): Rest-frame wavelength range. Col.(4): Estimated error and its standard
deviation. Col.(5): Velocity range . Col.(6): Velocity range width.}
\tablenotetext{*}{Used redshift is z=0.2972. Central peak wavelength is 6327 \AA \, in the observed frame 
(4877.5 \AA \, in the rest-frame). Shift relative to the narrow H$\beta$ is 21 \AA \, or 1013 km s$^{-1}$ in the observed frame (16.5 \AA \, or 1018 km s$^{-1}$ in the rest-frame).}
\end{deluxetable}
%%%%%%%%%%%%%%%%%%%%%%%%%%%%%%%%

\clearpage

%Table Segment Fluxes
%%%%%%%%%%%%%%%%%%%%%%%%%%%%%%%%
\begin{deluxetable}{clccccc}
\tabletypesize{\scriptsize}
\tablecaption{The measured broad-line and line-segment fluxes of H$\beta$
in units of $10^{-13} \rm erg \, cm^{-2} s^{-1} $.\label{segm}}
\tablewidth{0pt}
\tablehead{
\colhead{N} & \colhead{UT-date} & \colhead{JD+} & \colhead{F(H$\beta$)broad} & \colhead{F(H$\beta$)core} &
\colhead{F(H$\beta$)red} & \colhead{F(H$\beta$)far red}  \\
  &    & \colhead{2400000+} &  & & &   \\
\colhead{1} & \colhead{2} & \colhead{3} & \colhead{4} & \colhead{5} & \colhead{6} & \colhead{7} 
}
\startdata
   1 & 48095.00  &7.171 $\pm$0.205&  0.861 $\pm$ 0.042  & 3.672 $\pm$ 0.114 & 1.838 $\pm$ 0.075   &  0.704 $\pm$ 0.042  \\
   2 & 48476.00  &6.979 $\pm$0.200&  0.822 $\pm$ 0.040  & 3.714 $\pm$ 0.115 & 1.731 $\pm$ 0.071   &  0.638 $\pm$ 0.038  \\
   3 & 48479.00  &6.427 $\pm$0.184&  0.673 $\pm$ 0.033  & 3.466 $\pm$ 0.107 & 1.563 $\pm$ 0.064   &  0.603 $\pm$ 0.036  \\
   4 & 48813.00  &7.235 $\pm$0.207&  0.763 $\pm$ 0.037  & 3.806 $\pm$ 0.118 & 1.910 $\pm$ 0.078   &  0.717 $\pm$ 0.043  \\
   5 & 48860.00  &7.702 $\pm$0.220&  0.805 $\pm$ 0.040  & 4.024 $\pm$ 0.125 & 1.960 $\pm$ 0.080   &  0.802 $\pm$ 0.048  \\
   6 & 49597.00  &7.060 $\pm$0.202&  0.763 $\pm$ 0.037  & 3.541 $\pm$ 0.110 & 1.917 $\pm$ 0.079   &  0.724 $\pm$ 0.043  \\
   7 & 50866.58  &8.024 $\pm$0.229&  0.842 $\pm$ 0.041  & 4.128 $\pm$ 0.128 & 2.131 $\pm$ 0.087   &  0.821 $\pm$ 0.049   \\
   8 & 50941.07  &7.740 $\pm$0.221&  0.783 $\pm$ 0.038  & 4.000 $\pm$ 0.124 & 2.076 $\pm$ 0.085   &  0.791 $\pm$ 0.047   \\
\enddata
\end{deluxetable}
%%%%%%%%%%%%%%%%%%%%%%%%%%%%%%%%

%Table  (ph_flares.txt).
%%%%%%%%%%%%%%%%%%%%%%%%%%%%%%%%
\begin{deluxetable}{lccccc}
\tabletypesize{\scriptsize}
%\rotate
\tablecaption{Detected flare-like events in photometric data.\label{flare}}
\tablewidth{0pt}
\tablehead{
 \colhead{N} & \colhead{UT-Date} &  \colhead{MJD} & \colhead{Amplitude} & \colhead{Imax/Imin} 
&\colhead{maxJD(i+1)-maxJD(i)} \\
  &      &  \colhead{2400000}  & \colhead{[m]} & & \colhead{[days]}\\
  }
\startdata
 1 & 2005Nov28  &  53703  &   0.54    &     1.6   &      1066 (2-1) \\
 2 & 2008Oct29  &  54769  &   0.10    &     1.1   &      1030 (3-2) \\ 
 3 & 2011Aug26  &  55799  &   0.17    &     1.17  &      1002 (4-3) \\
 4 & 2014May23  &  56801  &   0.10    &     1.1   &   \\
\enddata
\tablecomments{Col.(1): number of flare-like event; Col.(2): UT-date; Col.(3): Modified Julian Date;
Col. (4): approximate amplitude between maximum and minimum (in magnitudes) for a flare-like event; 
Col.(5): the ratio of the maximum to minimum intensity; 
Col.(6): the difference in days between the two relevant events (in parentheses).}
\end{deluxetable}

%Variability parameters
%%%%%%%%%%%%%%%%%%%%%%%%%%%%%%%%
\begin{deluxetable}{lccccc}
%\tabletypesize{\scriptsize}
\tablecaption{Parameters of the continuum and line variabilities.\label{var}}
\tablewidth{0pt}
\tablehead{
 \colhead{Feature}  &  \colhead{N}    &  \colhead{$F$(mean)\tablenotemark{a}}  &  
\colhead{$\sigma$($F$)\tablenotemark{a}} & \colhead{$R$(max/min)} & \colhead{$F$(var)}\\
\colhead{1} & \colhead{2} & \colhead{3} & \colhead{4} & \colhead{5} & \colhead{6} }
\startdata
 cont 5100    & 127 &    7.14    & 1.36   &  2.17    &  0.189 \\
 cont 4200    & 76  &    10.41   & 2.00   &  2.14    &  0.189 \\
 H$\beta$ - total  & 127 &    12.45   & 0.95   &  1.41    &  0.071 \\
 H$\gamma$ - total & 76  &    2.89    & 0.57   &  2.14    &  0.190 \\
 H$\beta$ - broad  & 127 &    8.01   & 0.86   & 1.75    &  0.103  \\
 H$\beta$ - blue   & 127 &    0.87   & 0.11   & 1.78    &  0.116 \\
 H$\beta$ - core   & 127 &    4.15   & 0.53   & 1.81    &  0.123 \\
 H$\beta$ - red    & 127 &    2.10   & 0.24   & 1.90    &  0.105 \\
 H$\beta$ - far red & 127&    0.78   & 0.10   & 2.02    &  0.106 \\ 
\enddata
\tablecomments{Col.(1): Analyzed feature of the spectrum. Col.(2): Total number of spectra. 
 Col.(3): Mean flux.\tablenotemark{a} Col.(4): Standard
deviation\tablenotemark{a}. Col.(5): Ratio of the maximal to minimal flux . Col.(6): Variation amplitude (see text).}
\tablenotetext{a}{Continuum flux is in units of $10^{-15} \rm erg \ cm^{-2} s^{-1} A^{-1}$.
and line fluxes and line-segment fluxes are in units of $10^{-13} \rm erg \ cm^{-2}s^{-1}$.} 
\end{deluxetable}
%%%%%%%%%%%%%%%%%%%%%%%%%%%%%%%%

%%%%%%%%%%%%%%%%%%%%%%%%%%%%%%%%
\begin{deluxetable}{lcccccccccc}
\tabletypesize{\scriptsize}
\tablecaption{Time-lag analysis of observed and Gausssian Procces Regression (GPR) surrogates time series (TS)
of continua, H$\beta$, and H$\gamma$ light curves.\label{ccf}}
\tablewidth{0pt}
\tablehead{
 \colhead{TS} &  \colhead{TS1} &  \colhead{TS2} &  \colhead{${\bar{P}}$} & \colhead{$\tilde{P}$} & \colhead{N}
 & \colhead{$\tau_{ ZDCF}$}& \colhead{$r_{ZDCF}$} & \colhead{$\tau_{DCF}$} & \colhead{$r_{DCF}$} & \colhead{$\tau_{SPEAR}$}\\
\colhead{1} & \colhead{2} & \colhead{3} & \colhead{4} & \colhead{5} & \colhead{6} & \colhead{7} & \colhead{8} & \colhead{9}
& \colhead{10}& \colhead{11} }
\startdata
\multirow{2}{*}{\rotatebox[origin=c]{90}{Observed}} &cnt5100 & H$\beta$&  48.7 &28.6 &121 &${26}_{-24}^{+48}$ &$0.59_{-0.07}^{+0.06}$ & ${208}_{-11}^{+11}$& $0.72_{-0.02}^{+0.02}$ & ${196}_{-4}^{+9}$   \\
%  & & &&& & ${137}_{-19}^{+54}$ &$0.49_{-0.10}^{+0.10}$ & & & \\
&cnt4200 & H$\gamma$&82.3 & 54.9&72 &$65_{-7}^{+24}$ & $0.78_{-0.08}^{+0.07}$ &$54_{-19}^{+19}$& $0.50_{-0.20}^{+0.16}$ & $60_{-11}^{+6}$    \\
%& H$\beta$ & H$\gamma$& 48.7/82.3\tablenotemark{*}&28.6/54.9\tablenotemark{*}&121/72 &${51}_{-66}^{+36}$&$0.57_{-0.01}^{+0.01}%$& ${58}_{-11}^{+11}$ & $0.47_{-0.10}^{+0.14}$ & $11_{-3}^{+53}$   \\   
%& cnt4200 & cnt5100 &82.3/48.7\tablenotemark{*}&54.9/28.6\tablenotemark{*}&72/121 & $7_{-7}^{+3}$ & $0.87_{-0.03}^{+0.03}$ &  $34_{-5}^{+5}$  &  $0.41_{-0.20}^{+0.16}$  & $6_{-2}^{+7}$\\ 
%&M$_{\rm V}$ & M$_{\rm R}$ &43.5&28.9&96 &  $2.2_{-1.2}^{+6.7}$    & $0.98_{-0.01}^{+0.01}$ & $1.7_{-0.07}^{+0.07}$ & $0.69_{-0.04}^{+0.04}$ & $2.6_{-1.9}^{+3.3}$  \\
%&M$_{\rm B}$ & M$_{\rm R}$ &43.5&28.9&96 & $2.1_{-1.1}^{+6.8}$ & $0.97_{-0.01}^{+0.01}$ & $1.7_{-0.1}^{+0.1}$ & $0.69_{-0.04}^{+0.04}$ & $1.9_{-1.3}^{+1.7}$\\ 
%&M$_{\rm B}$ & M$_{\rm V}$ &43.5&28.9&96 & $2.1_{-1.1}^{+6.7}$ & $0.96_{-0.02}^{+0.01}$ & $1.7_{-0.1}^{+0.1}$ & $0.69_{-0.04}^{+0.04}$ & $2.5_{-1.6}^{+1.6}$\\ 
&&&&&&&&\\
\hline
\multirow{2}{*}{\rotatebox[origin=c]{90}{GPR}} 
&cnt5100 & H$\beta$&  11.7&11.7&500 &${118.0}_{-0.002}^{+0.100}$ &$0.73_{-0.05}^{+0.05}$ & ${125.0}_{-0.3}^{+0.3}$ &$0.96_{-0.003}^{+0.003}$ & ${120.6}_{-7.5}^{+10.6}$  \\
&cnt4200 & H$\gamma$&11.7&11.7&500 & $152.3_{-0.001}^{+0.001}$ & $0.90_{-0.01}^{+0.01}$ & 
 $96.0_{-4.0}^{+4.0}$ & $0.90_{-0.02}^{+0.02}$  &  $59.3_{-1.9}^{+0.6}$   \\
\enddata
\tablecomments{Col.(1) - TS type; Cols. (2) and (3) cross-correlated TS; Cols.(4) and (5) mean ${ \bar{P}}$ and median $\tilde{P}$ sampling; Col.(6) number of points in TS; Col.(7): time-lag calculated using ZDCF method; Col.(8): cross correlation coefficient calculated using ZDCF; Col.(9): time-lag calculated using DCF; Col. (10): cross correlation coefficient
calculated using DCF; Col. (11) time-lag calculated using SPEAR method.}
%\tablenotetext{*}{TS1 value/TS2 value is given in the case of different TS length.}
\end{deluxetable}
%%%%%%%%%%%%%%%%%%%%%%%%%%%%%%%%


\begin{thebibliography}{}

%
\bibitem[\protect\citeauthoryear{Alexander}{1997}]{A97} Alexander, T., ZDCF Analysis of Small Samples of Sparse Light Curves, 
Astronomical Time Series, Eds. D. Maoz, A. Sternberg, and E.M. Leibowitz, 1997 (Dordrecht: Kluwer), p. 163-166, 1997.

%
\bibitem[\protect\citeauthoryear{Alexander}{2013}]{A13} Alexander, T, 2013, http://adsabs.harvard.edu/abs/2013arXiv1302.1508A

%
\bibitem[\protect\citeauthoryear{Amirkhanian et al.}{2000}]{am00} Amirkhanian V., et al., 2000, Bull. Spec. Astrophys. Obs. 50

%
\bibitem[\protect\citeauthoryear{Aravena et al.}{2011}]{ar11}
	Aravena, M., Wagg, J., Papadopoulos, P. P., Feain, I. J., 2011, ApJ, 737, 64.

%
\bibitem[\protect\citeauthoryear{Bahcall et al.}{1992}]{ba92}
Bahcall, J. N., Jannuzi, B. T., Schneider, D. P., Hartig, G. F., Green, R. F., 1992, ApJ, 397, 68


\bibitem[\protect\citeauthoryear{Bentz et al.}{2010}]{be10} Bentz, M.C., Walsh, J.L., Barth, A.J. et al.
ApJ, 716, 993
%
\bibitem[\protect\citeauthoryear{Blundell et al.}{1996}]{bl96}
Blundell, K. M., Beasley, A. J., Lacy, M., Garrington, S. T., 1996, ApJ, 468L, 91

%
\bibitem[\protect\citeauthoryear{Blundell \& Lacy}{1995}]{bl95}
Blundell, K. M.,  Lacy, M., 1995, MNRAS, 274L, 9

%
\bibitem[\protect\citeauthoryear{Blundell \& Rawlings}{2001}]{bl01}
Blundell, K. M., Rawlings, S., 2001, ApJ, 562L, 5


%
\bibitem[\protect\citeauthoryear{Bon et al.}{2012}]{bo12}
Bon, E., Jovanovi\'c, P., Marziani, P., Shapovalova, A. I., Bon, N., Borka Jovanovi\'c, V.,
Borka, D., Sulentic, J., Popovi\'c, L. \v C., 2012, ApJ, 759, 118

%
\bibitem[\protect\citeauthoryear{Bogdanovi\'c}{2015}]{bo15}
Bogdanovi\'c, T., 2015, Gravitational Wave Astrophysics, Astrophysics and Space Science Proceedings,  40, 115.


%
\bibitem[\protect\citeauthoryear{Cousins}{1976}]{co76} Cousins, A.W.J., 1976, Mem.R.Astron.Soc. 81, 25

\bibitem[\protect\citeauthoryear{Cuadra et al.}{2009}]{cu09}
 Cuadra, J., Armitage, P.J., Alexander, R.D., Begelman, M.C. 2009,
MNRAS, 393, 1423

 \bibitem[\protect\citeauthoryear{Denney et al.}{2009}]{de09} Denney, K.~D., Watson, 
L.~C., Peterson, B.~M., et al.\ 2009, \apj, 702, 1353 

 \bibitem[\protect\citeauthoryear{Dietrich et al.}{1998}]{di98}  Dietrich, M., 
Peterson, B.~M., Albrecht, P., et al.\ 1998, \apjs, 115, 185

%
%
\bibitem[\protect\citeauthoryear{Dimitrijevi\'c et al.}{2007}]{dim07} Dimitrijevi\'c, M. S., Popovi\'c,
 L. \v C., Kova\v cevi\'c, J., Da\v ci\'c, M., Ili\'c, D. 2007, MNRAS, 374,1181

\bibitem[\protect\citeauthoryear{Edelson \& Krolik}{1988}]{EK188} 
Edelson, R. A., \& Krolik, J. H., ApJ, 333, 646-659, 1988

\bibitem[\protect\citeauthoryear{Gaskell}{2009}]{ga09} Gaskell, C. M. 2009, NewAR, 53, 140103

\bibitem[\protect\citeauthoryear{Floyd et al.}{2004}]{fl04}
Floyd, D. J. E., Kukula, M. J., Dunlop, J. S.
et al. 2004, MNRAS, 355, 196
%\bibitem[\protect\citeauthoryear{Fromerth \& Melia}{2000}] {FM00} Fromerth, M, J, Melia, F,2000, \apj, 533, 172


%	
\bibitem[\protect\citeauthoryear{Fried}{1998}]{fr98}
Fried, J. W., 1998, A\&A, 331L, 73
	

%
\bibitem[\protect\citeauthoryear{Hutchings \& Neff}{1991}]{hn91}
Hutchings, J. B., Neff, S. G., 1991, AJ, 101, 2001

%
\bibitem[\protect\citeauthoryear{Ivezi\'c et al.}{2014}]{iv14} Ivezi\'c, \v Z., Connolly, A.J., 
Vanderplas, J.T., Gray, A., 2014, Statistics, Data Mining and Machine Learning in Astronomy, 
Princeton University Press, Princeton, NJ


%
\bibitem[\protect\citeauthoryear{Kelly et al.}{2009}]{k09} Kelly, B.C., Bechtold, J., \& Siemiginowska, A., 2009, ApJ, 698, 895
%
\bibitem[\protect\citeauthoryear{Kollatschny et al.}{2006}]{ko06}
Kollatschny, W.; Zetzl, M.; Dietrich, M., 2006, A\&A, 454, 459

%
\bibitem[\protect\citeauthoryear{Kolman et al.}{1993}]{ko93}
Kolman, M., Halpern, J. P., Shrader, C. R., Filippenko, A. V., Fink, H. H., Schaeidt, S. G., 1993, ApJ, 402, 514

%
\bibitem[\protect\citeauthoryear{Kolman et al.}{1991}]{ko91}
Kolman, M., Halpern, J. P., Shrader, C. R., Filippenko, A. V., 1991, ApJ, 373, 57


\bibitem[\protect\citeauthoryear{Korista et al.}{1997}]{ko97} 	Korista, K., Baldwin, J., Ferland, G.,  Verner, D. 
1997, ApJS 108, 401
\bibitem[\protect\citeauthoryear{Korista \& Goad}{2004}]{kg04}
Korista, K. T., Goad, M. R. 2004, ApJ, 606, 749, (erratum 627, 577 [2005])

%
\bibitem[\protect\citeauthoryear{Kova\v cevi\'c et al.}{2010}]{ko10}
	Kova\v cevi\'c, J., Popovi\'c L. \v C., Dimitrijevi\'c, M.S. 2010, ApJS, 189, 15

\bibitem[\protect\citeauthoryear{Kova\v cevi\'c et al.}{2014}]{ko14} Kova{\v 
c}evi{\'c}, A., Popovi{\'c}, L.~{\v C}., Shapovalova, A.~I., et al.\ 2014, 
Advances in Space Research, 54, 1414 

%
\bibitem[\protect\citeauthoryear{Kozlowski et al.}{2010}]{koz10}
Kozlowski, S., Kochanek, C.S., Udalski, A., Wyrzykowski, L., Soszynski, I., et al., 2010, ApJ, 708, 927

%
\bibitem[\protect\citeauthoryear{Lacy et al.}{1992}]{la92}
Lacy, M., Rawlings, S., Hill, G. J., 1992, MNRAS, 258, 828


%	
\bibitem[\protect\citeauthoryear{Landt et al.}{2008}]{la08}
Landt, H., Bentz, M. C., Ward, M. J., Elvis, M., Peterson, B. M., Korista, K. T., Karovska, M., 2008, ApJS, 174, 282


\bibitem[\protect\citeauthoryear{Lomb}{1976}]{l76} Lomb, N.R., 1976, \apss, 39,  447

\bibitem[\protect\citeauthoryear{Lousto et al.}{2010}]{lo10}  Lousto, C.O.,  Nakano, H., Zlochower, Y.,  Campanelli  M. 2010, PhRvD,  81h4023
%
\bibitem[\protect\citeauthoryear{McLeod et al.}{2010}]{ma10}
MacLeod, C.L., Ivezi\'c, \v Z., Kochanek, C.S., KozLowski, S., Kelly, B., et al., 2010, ApJ, 721, 1014

%
\bibitem[\protect\citeauthoryear{McLeod \& McLeod}{2001}]{mm01}
McLeod, K. K.; McLeod, B. A., 2001, ApJ, 546, 782

%
\bibitem[\protect\citeauthoryear{Mortier et al.}{2015}]{mo15} Mortier, A., Faria, J.~P., Correia, C.~M., Santerne, A., \& Santos, N.~C.\ 2015, \aap, 573, A101


%
\bibitem[\protect\citeauthoryear{O'Brien et al.}{1998}]{ob98} O'Brien, P.T., Dietrich, M., Leighly, K. et al. 1998, \apj, 509, 163

\bibitem[\protect\citeauthoryear{Oegerle et al.}{2000}]{oe00} Oegerle, W. R., Tripp, T. M., Sembach, K. R. et al. 2000, \apjl, 538, 23L

\bibitem[\protect\citeauthoryear{Onken et al.}{2004}]{on04}
Onken, C. A., Ferrarese, L., Merritt, D., Peterson, B. M., Pogge, R. W., Vestergaard, M., Wandel, A., 2004, ApJ, 615, 645
%

\bibitem[\protect\citeauthoryear{Penston et al.}{1971}]{pen71}  Penston, M.J., Penston, M.V., Sandage, A., 1971, PASP 83, 783
%
\bibitem[\protect\citeauthoryear{Peterson}{2014}]{pe14}  Peterson, B.~M., 2014, SSRv,  183,  253

%
\bibitem[\protect\citeauthoryear{Peterson}{1993}]{pet93}  Peterson, B. M. 1993, \pasp, 105, 247


%
\bibitem[\protect\citeauthoryear{Peterson et al.}{1994}]{pet94} Peterson, B.~M.,
Berlind, P., Bertram, R., et al.\ 1994, \apj, 425, 622

%
\bibitem[\protect\citeauthoryear{Peterson et al.}{2002}]{pet02} Peterson, B.~M.,
Berlind, P., Bertram, R., et al.\ 2002, \apj, 581, 197

%
\bibitem[\protect\citeauthoryear{Peterson \& Collins}{1983}]{pc83} Peterson, B.M., \& Collins II, G.W. 1983, \apj, 270, 71

%
\bibitem[\protect\citeauthoryear{Peterson et al.}{1995}]{pet95} Peterson, B.~M., Pogge, R.~W., Wanders, I., Smith, S.~M.,
\& Romanishin, W.\ 1995, \pasp, 107, 579
%
\bibitem[\protect\citeauthoryear{Peterson et al.}{1998}] {pet98} Peterson, B.~M.,
Wanders, I., Bertram, R., et al. 1998, \apj, 501, 82

%
\bibitem[\protect\citeauthoryear{Popovi\'c}{2012}]{pop12} Popovi\'c, L. \v C., 2012, NewAR, 56, 74

%
\bibitem[\protect\citeauthoryear{Popovi\'c et al.}{2004}]{pop04}
Popovi\'c, L. \v C., Mediavilla, E., Bon, E., \& Ili\'c, D. 2004
A\&A 423, 909

%
\bibitem[\protect\citeauthoryear{Popovi\'c et al.}{2011}]{pop11} Popovi\'c, L. \v C., Shapovalova, A. I., Ili\'c, D. et al.
2011, A\&A, 528A, 130
%
\bibitem[\protect\citeauthoryear{Pravdo \& Marshall}{1984}]{pm84}
Pravdo, S. H., Marshall, F. E., 1984, ApJ, 281, 570

%
\bibitem[\protect\citeauthoryear{Rasmussen \& Williams}{2006}]{EK88} Rasmussen, C. E. and  Williams, C., Gaussian 
Processes for Machine Learning, the MIT Press 2006
   
%
\bibitem[\protect\citeauthoryear{Rehfeld et al.}{2011}]{re11} Rehfeld,K, Marwan, N,  Heitzig, J and  Kurths, J., Nonlin.,
   Processes Geophys., 18, 389–404, 2011

%
\bibitem[\protect\citeauthoryear{Reynolds et al.}{2014}]{re14}
Reynolds, C. S., Lohfink, A. M., Babul, A., Fabian, A. C., Hlavacek-Larrondo, J., Russell, H. R., Walker, S. A.,
2014, ApJ, 792L, 41
   
 \bibitem[\protect\citeauthoryear{Robinson et al.}{2010}]{ro10}  
	Robinson, A., Young, S., Axon, D. J., Kharb, P., Smith, J. E., 2010, ApJ, 717L, 122
%
   \bibitem[\protect\citeauthoryear{Russell et al.}{2010}]{ru10}
Russell, H. R., Fabian, A. C., Sanders, J. S., Johnstone, R. M., Blundell, K. M., Brandt, W. N., Crawford, C. S.,
2010, MNRAS, 402, 1561
%\bibitem[\protect\citeauthoryear{Salpeter}{1964}]{s64} Salpeter, E. E. 1964, ApJ, 140, 796

\bibitem[\protect\citeauthoryear{Scargle}{1982}]{sc82} Scargle, J.D., 1982, \apj,  263,  835

\bibitem[\protect\citeauthoryear{Schmitt et al.}{2003}]{sc03}	
Schmitt, H. R., Donley, J. L., Antonucci, R. R. J., Hutchings, J. B., Kinney, A. L., Pringle, J. E., 2003, ApJ, 597, 768

\bibitem[\protect\citeauthoryear{Schneider et al.}{1992}]{sc92}
Schneider, D. P., Bahcall, J. N., Gunn, J. E., Dressler, A., 1992, AJ, 103, 1047

%
\bibitem[\protect\citeauthoryear{Shapovalova et al.}{2001}]{sh01} Shapovalova, A.~I., Burenkov,
A.~N., Carrasco, L., et al.\ 2001, \aap, 376, 775

%
\bibitem[\protect\citeauthoryear{Shapovalova et al.}{2004}]{sh04} Shapovalova, A.I., Doroshenko, V.T.,
Bochkarev, N.G, et al. 2004, \aap, 422, 925

%
\bibitem[\protect\citeauthoryear{Shapovalova et al.}{2010}]{sh10} Shapovalova, A. I., 
Popovi\'c, L.\v C., Bochkarev, N.G., et al. 2010, \aap, 517A, 42

%
\bibitem[\protect\citeauthoryear{Shapovalova et al.}{2012}]{sh12} Shapovalova, A.I., Popovi\'c, L.\v C., Burenkov, A. N., 
et al. \ 2012, \apjs, 202, 10

%
\bibitem[\protect\citeauthoryear{Shapovalova et al.}{2013}]{sh13} Shapovalova, A. I., Popovi\'c, L.\v C., Bochkarev, N.G., et al.,
2013, A\&A, 559A, 10S

%
\bibitem[\protect\citeauthoryear{Sulentic et al.}{2000}]{su00} Sulentic, J. W., Marziani, P., Dultzin-Hacyan, D.2000, ARA\&A, 38, 521

\bibitem[\protect\citeauthoryear{Ulrich et al.}{1992}]{ul92}
Ulrich, M.-H., Fink, H. H., Schaeidt, S., Baganoff, F., Malkan, M. A., Heidt, J., Wagner, S., 1992, A\&A, 266, 183
%
\bibitem[\protect\citeauthoryear{Vanderplas et al.}{2012}]{va12} Vanderplas, J.T., Connolly, A.J., 
Ivezi\'c, \v Z., Gray, A., 2012, Conference on Intelligent Data Understanding (CIDU), Introduction 
to astroML: Machine learning for astrophysics, oct, 47 -54

%
\bibitem[\protect\citeauthoryear{van Groningen \& Wanders}{1992}]{vg92} Van Groningen, E. \& Wanders, I.\ 1992, PASP, 104, 700

%
\bibitem[\protect\citeauthoryear{Vlasyuk}{1993}] {vl93} Vlasyuk V.V., 1993, Bull. Spec. Astrophys. Obs. 36, 107
	
%
\bibitem[\protect\citeauthoryear{Walker et. al.}{2014}] {wa14}	
Walker, S. A., Fabian, A. C., Russell, H. R., Sanders, J. S., 2014, MNRAS, 442, 2809

\bibitem[\protect\citeauthoryear{Wandel et al.}{1999}] {wa99} Wandel, A., Peterson,
B.~M., \& Malkan, M.~A.\ 1999, \apj, 526, 579

%\bibitem[\protect\citeauthoryear{Wu \& Liu}{2004}]{wu04} Wu, X.-B., Liu, F. K. 2004, ApJ, 614, 91
\bibitem[\protect\citeauthoryear{Zanotti et al.}{2010}]{za10}
 Zanotti, O.,  Rezzolla, L., Del Zanna, L.,  Palenzuela C. 2010,
A\&A, 523A, 8
%
\bibitem[\protect\citeauthoryear{Zu et al.}{2011}]{zu11} Zu, Y., Kochanek, C. S., Peterson, Bradley M. 2011, ApJ, 735, 80.

\end{thebibliography}
\end{document}